\documentclass[twocolumn,seceq]{jpsj2} %% two-column layout
\def\Vec#1{\mbox{\boldmath $#1$}}
\def\kv{\mbox{\boldmath $ k$}}

\def\rv{\mbox{\boldmath $r$}}
\def\cg{\check{g}}

\def\cP{\check{P}}
\def\ha{\hat{a}}
\def\hb{\hat{b}}

\def\ab{(1+ \ha \hb)^{-1}}
\def\ba{(1+ \hb \ha)^{-1}}

\def\hDelta{\hat{\Delta}}

\title{Analytical Formulation of the Local Density of States around a Vortex Core in Unconventional Superconductors}

\author{\textsc{Yuki Nagai}$^{1}$\thanks{E-mail address: ss56079@mail.ecc.u-tokyo.ac.jp}, \textsc{Yosuke Ueno}$^{2}$, 
 \textsc{Yusuke Kato}$^{3}$, and \textsc{Nobuhiko Hayashi}$^{4}$}

\inst{$^{1}$Department of Physics, University of Tokyo, Tokyo 113-0033, Japan\\
$^{2}$Department of Applied Physics, University of Tokyo, Tokyo 113-8656, Japan\\
$^{3}$Department of Basic Science, University of Tokyo, Tokyo 153-8902, Japan\\
$^{4}$Institut f$\ddot{u}$r Theoretische Physik, ETH-Z$\ddot{u}$rich, CH-8093 Z$\ddot{u}$rich, Switzerland}

\abst{
On the basis of the quasiclassical theory of superconductivity, we obtain a formula for the local density of states (LDOS) around a 
vortex core of superconductors with anisotropic pair-potential and Fermi surface in arbitrary directions of magnetic fields.
Earlier results on the LDOS of $d$-wave superconductors and NbSe$_2$ are naturally interpreted within our theory 
geometrically; the region with high intensity of the LDOS observed in numerical calculations turns out to the enveloping 
curve of the trajectory of Andreev bound states. 
We discuss experimental 
results on YNi$_2$B$_2$C within the quasiclassical theory of superconductivity.

}

\kword{Unconvensional superconductivity, Vortex core, Local density of states, Quasiclassical theory}

\begin{document}
\maketitle

\section{Introduction} %% No sections necessary for express letters, letters and short notes
One of the aims of the study on the condensed matter is to identify 
the order parameter in a symmetry-broken state. In supercoducting materials, 
the order parameter is given by the pairing wavefunction, which is related to 
the pair-potential. 
As experimental techniques to observe the properties of the pair-potential, 
several probes are available such as nuclear-spin-lattice relaxation, the measurement of specific heat, (angular-resolved) 
photo-emission spectroscopy, field direction dependence of thermal conductivity and so on. Among 
those probes, we focus particularly on the scanning tunneling microscopy and spectroscopy (STM/STS), which is now a 
main technology in nano-science. 

The observation of the local density of states (LDOS) in the vortex phase in superconductors yields 
 rich information on the pairing symmetry; the LDOS around a vortex core mainly comes from Andreev 
bound states, whose spectrum is sensitive to the anisotropy of the pair-potential. 
Thus the anisotropic spatial 
pattern of the LDOS around a vortex core is expected to be observed as a consequence of the anisotropy 
of pair-potential on the Fermi surface. 
Indeed the LDOS of NbSe$_2$\cite{Hess} and YNi$_2$B$_2$C\cite{Nishimori} measured 
by STM reveal, respectively, sixfold star-shaped and fourfold structure.  Motivated by the experimental 
results on NbSe$_2$, Gygi and Schl\" uter\cite{Gygi91} and Hayashi {\it et al}. \cite{hayashi} successfully 
reproduced the LDOS around a vortex by numerical calculations. The LDOS around a vortex in non-$s$-wave 
superconductors was firstly obtained by Schopohl and Maki\cite{Schopohl2} in a numerical study of $d$-wave 
vortex with the use of Riccati formalism of quasiclassical theory. 
Ichioka {\it et al.},\cite{ichioka} subsequently, obtained the LDOS 
of $d$-wave vortex in the self-consistent numerical calculations. 
Prior to ref.~\citen{Schopohl2}, Riccati formalism 
was developed by Ashida {\it et al.} \cite{Ashida89} in the context of boundary problem at Supercondutor-Normal-metal 
interfaces. \cite{nagato}
The formalisms of refs.~\citen{Ashida89} and \citen{Schopohl3} turned out to be equivalent although it is not 
so obvious. 
Riccati formalism was extended to non-equilibrium case by Eschrig\cite{eschrig} via the 
projection operator formalism by Shelankov\cite{Shelankov}. 

Although there exist many numerical works on the LDOS, each calculation usually takes much machine-time. 
It would be elaborate to treat three-dimensional superconductors with anisotropic Fermi surface and 
pairing interactions. 
Further, the LDOS of refs.~\citen{hayashi} and \citen{Schopohl2} has strong spatial dependence 
and hence a comprehensive interpretation by an analytical theory is highly desirable. The bound states 
localized near an isotropic $s$-wave vortex core were studied analytically by Caroli {\it et al.} \cite{Caroli} with 
the use of Bogoliubov-de Gennes equation. Equivalently, an approximate expression for the quasiclassical Green 
function near a vortex core for low energy was analytically obtained by Kramer and Pesch.\cite{Kramer} 
We call their 
approximation ^^ ^^ Kramer-Pesch approximation(KPA)" in the present paper. 
Although the spirit of KPA is 
hard to understand as it stands, it becomes clear after transforming quasiclassical equation into Riccati 
formalism; KPA is equivalent to first-order perturbation with respect to energy and impact parameter. 
KPA was successfully applied to two-dimensional $d$-wave superconductors\cite{Schopohl2}, 
in a slightly different formalism. \cite{Waxman}

In this paper, we apply KPA to more general case; the LDOS of two-dimensional and three-dimensional 
superconductors with anisotropic Fermi surface and pair-potential for arbitrary direction of 
magnetic fields are discussed. 
Our theory yields a natural geometrical interpretation of spatially 
anisotropic pattern of the LDOS for anisotropic superconductors. 
We will see that earlier numerical 
results on the LDOS are easily understood within our theory. 
Furthermore, we discuss experimental 
results on YNi$_2$B$_2$C\cite{Nishimori}. 
Contrary to the claim of the authors of ref.~\citen{Nishimori}, 
we show that a part of the experimental results is naturally interpreted by  the quasiclassical theory. 
In an earlier short report, the application of our theory to CePt$_3$Si was already reported\cite{nagai}. 
One of the aims of this paper is to present the details of the derivation. 
The unpublished results of ref.~\citen{Ueno} on 
two-dimensional superconductors are contained as a part of this paper (\S 3.1.2). 

This paper is organized as follows. 
In \S 2, we summarize the formulation of the quasiclassical theory of superconductivity.
In \S 3, we present our analytical formulation for the LDOS around a vortex core in the case of the unitary pair-potential 
($\hDelta \hDelta^{\dagger} \propto \hat{\sigma}_0$) of the spin singlet or the spin triplet, 
using the Riccati formalism and Kramer-Pesch approximation.
In \S 4, we calculate the specific LDOS patterns about NbSe$_2$ and YNi$_2$B$_2$C with our formulation.
In \S 5, some remarks on the formulation are given.
In \S 6, we compare our results with those of STM measurements on YNi$_2$B$_2$C.
In \S 7, the summary is given.
In the Appendix A, we describe the formulation to obtain the matrix Riccati equation.
In the Appendix B, we calculate the LDOS of CePt$_3$Si as an example of the case of the pair-potential which does not satisfy 
the relation $\hat{\Delta} \hat{\Delta}^{\dagger} \propto \hat{\sigma}_0$.
%%%%%%%%%%%%%%%%%%%%%%%%%%%%%%%%%%%%%%%%%%%%%%%%%%%%%%%%%
\section{Quasiclassical Theory of Superconductivity}
%%%%%%%5
We consider the pair-potential:
%%%%%%%%%%%%%%%%%%%%
\begin{eqnarray}
\Delta_{\eta \eta'}(\Vec{R}, \Vec{r}) &=& \Bigl{\langle} \psi_{\eta} \left( \Vec{R}+\frac{\Vec{r}}{2} \right) 
\psi_{\eta'} \left( \Vec{R}-\frac{\Vec{r}}{2} \right)
\Bigl{\rangle}, \\
\Delta_{\eta \eta'} (\Vec{R}, \kv) &=& \int d \rv e^{- {\rm i} \kv \cdot \rv} \Delta_{\eta \eta'} (\Vec{R}, \rv),
\end{eqnarray}
%%%%%%%%%%%%%%
where $\Vec{r}$ and $\Vec{R}$ are the relative and center-of-mass coodinates, respectively.
Here, $\psi_{\eta}(\rv)$ is the field operator for the quasiparticle state with spin $\eta$ 
and we use units in which $\hbar = k_{\rm B} = 1$.
We assume the weak coupling interaction so that the pair-potential is not zero only near the Fermi surface.
We consider pair-potential written as either $\hat{\Delta}(\Vec{R},\kv) = {\rm i} \psi(\kv) \hat{\sigma}_y A(\Vec{R})$ for singlet pairing or 
$\hat{\Delta}(\Vec{R}, \kv) = {\rm i} (\Vec{d}(\kv) \cdot \hat{\Vec{\sigma}} ) \hat{\sigma}_y A(\Vec{R})$ for triplet pairing.
Here $A(\Vec{R})$ is a function of $\Vec{R}$. 
If the coherence length $\xi$ is large compared to the Fermi wave 
length ($\sim 1/ k_{\rm F}$), 
we can calculate the LDOS around a vortex core on the basis of the quasiclassical theory of superconductivity.\cite{Eilen,Serene,Larkin}
 We consider the quasiclassical Green function $\check{g}$ that has the matrix elements in the Nambu (particle-hole) space as
\begin{eqnarray}
\check{g}(\rv,\tilde{\Vec{k}}, {\rm i} \omega_n) =  
\left(
\begin{array}{cc}
\hat{g} &  \hat{f} \\
-  \hat{\bar{f}} &  \hat{\bar{g}}
\end{array}
\right)
,
\end{eqnarray}
where $\omega_n = \pi T(2 n +1)$ is the Matsubara frequency with the temperature $T$ and an integer $n$.
Here $\tilde{\Vec{k}}$ is a unit vector ($\tilde{\Vec{k}} = \Vec{k}/k_{\rm F}$).
Throughout the paper, ^^ ^^ \textit{hat}" $\hat{a}$ denotes the $2 \times 2$ matrix in the spin space, and 
^^ ^^ \textit{check}" $\check{a}$ denotes a $4 \times 4$ matrix composed of the $2 \times 2$ 
Nambu space and the $2 \times 2$ spin space\cite{hayashi_con2}.
The equation of motion for $\check{g}$ called Eilenberger equation is written as \cite{Eilen,Larkin}
\begin{equation}
{\rm i} \Vec{v}_{\rm F}(\Vec{\tilde{\kv}}) \cdot \nabla \check{g} + 
[{\rm i} \omega_n \check{\tau}_3  - \check{\Delta},\check{g}]=0,
\label{eq:eilen}
\end{equation}
where
\begin{eqnarray}
\check{\tau}_3 = 
\left(
\begin{array}{cc}
\hat{\sigma}_0 & 0 \\
0 & - \hat{\sigma}_0
\end{array}
\right),
\: \: \check{\Delta} = \left(
\begin{array}{cc}
0 & \hat{\Delta} \\
- \hat{\Delta}^{\dagger} & 0
\end{array}
\right).
\end{eqnarray}
Here $\Vec{v}_{\rm F}(\Vec{\tilde{\kv}})$ is the Fermi velocity. 
The commutator
$[\check{a}, \check{b}] = \check{a} \check{b} - \check{b} \check{a}$. 
The Green function $\check{g}$ satisfies the normalization condition: $\check{g}^2 = \check{1}$, where $\check{1}$ is the $4 \times 4$ unit matrix. 
Considering clean superconductors in the type II limit, we neglect the impurity effect and the vector potential.

%%%%%%%%%%%%%%%%%%%%%%%%%%%%%
The local density of states with the isotropic Fermi surface is given by 
%%%%%%%%%%%%%%%%%
\begin{equation}
\nu(\rv,\epsilon) = - \nu(0) \int \frac{d \Omega_k}{ 4 \pi} {\rm Re} \; {\rm tr} (\hat{g}^{\rm R}). \label{eq:ldos}
\end{equation}
%%%%%%%%%%%%%%%%%
Here, $\hat{g}^{\rm R}$ is retarded Green function and $\nu(0)$ denotes the density of states on Fermi surface 
in the normal metalic state.
The local density of states with the anisotropic Fermi surface is given by 
%%%%%%%%%%%%%%%%%%
\begin{equation}
\nu(\rv,\epsilon) = - \int \frac{d S_{\rm F} }{2 \pi^2 v_{\rm F}} 
 {\rm Re} \: {\rm tr} (\hat{g}^{\rm R}). \label{eq:ldosa}
\end{equation}
%%%%%%%%%%%%%%
Here, $d S_{\rm F}$ is the Fermi-surface area element and 
$v_{\rm F}$ is the modulus of Fermi velocity.
%%%%%%%%%%%%%%%%%%%%%%%%%%5
\section{Riccati Formalism and Kramer-Pesch Approximation}
%%%%%%%%%%%%%%%%%%%%%%%%%%%%%%%%55
In order to solve the quasiclassical equation (\ref{eq:eilen}), 
we simplify the equation with a parametrization for the propagators that 
satisfy the normalization condition.
Projection operators are defined by $\check{P}_{\pm} = \frac{1}{2} (\check{1} \mp \check{g})$.\cite{Shelankov,eschrig}
Using this projection method (See, Appendix A), we obtain matrix Riccati equations:
\begin{eqnarray}
  \Vec{v}_{\rm F} \cdot \Vec{\nabla} \ha + 2  \omega_n \ha + \ha \hat{\Delta}^{\dagger} \ha     -
  \hat{\Delta} &=& 0,
  \label{eq:a}\\
 \Vec{v}_{\rm F} \cdot \Vec{\nabla} \hb - 2  \omega_n  \hb - \hb \hat{\Delta} \hb + \hat{\Delta}^{\dagger} 
&=& 0
\label{eq:b},
\end{eqnarray}
where
\begin{eqnarray}
\cg 
= -
\check{N}
\left(
\begin{array}{cc}
(\hat{1} - \ha   \hb) & 2{\rm i} \ha   \\
- 2 {\rm i}\hb & - (\hat{1} - \hb \ha  )
\end{array}
\right),\label{eq:green}\\
\check{N} = \left(
\begin{array}{cc}
\ab & 0 \\
0 & \ba
\end{array}
\right).
\end{eqnarray}
Since these matrix Riccati equations (\ref{eq:a}) and (\ref{eq:b}) contain $\Vec{\nabla}$ 
only through $\Vec{v}_{\rm F} \cdot \Vec{\nabla}$, 
they reduce to a one-dimensional problem on a straight line, 
the direction of which is given by that of the Fermi velocity.
%%%%%%%%%%%%%%%%%%%%%%%%%%%%%%%%%%%%%%%%%%%%
\subsection{Two-dimensional Fermi surface}
\subsubsection{Frame of the space}
\begin{figure}[htbp]
 \begin{center}
  \includegraphics[width=6cm,keepaspectratio]{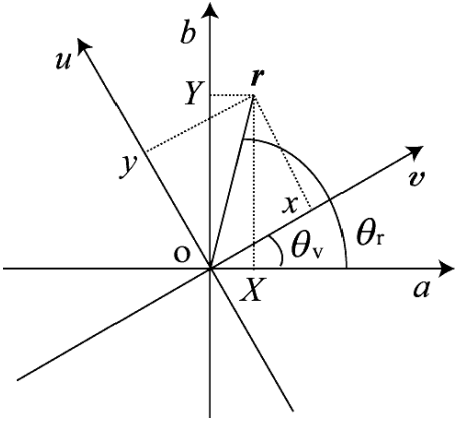}
\caption{Frame of the space. The direction of $\Vec{v}$ is equal to that of Fermi velocity.
$a$ and $b$ are the axes fixed to the crystal axes.
$\theta_r$ is the angle between the $a$ axis and $\Vec{r}$.}
\label{fig:za}
 \end{center}
\end{figure}

We consider the system with two-dimensional Fermi surface.
We introduce the frame of the space written as 
\begin{eqnarray}
\kv &=& k  \Bigl{[} \cos \theta \hat{\Vec{a}} + \sqrt{\frac{m_y}{m_x}} \sin \theta  \hat{\Vec{b}} \Bigl{]},  \\
\Vec{v} &=& v \Bigl{[} \cos \theta  \hat{\Vec{a}} + 
\sqrt{ \frac{m_x}{m_y} } \sin \theta  \hat{\Vec{b}}  \Bigl{]}, \label{eq:vtheta} \\
 &\equiv&  v_{\rm F}(\theta_v) \Bigl{[} \cos \theta_v \hat{\Vec{a}} 
 +  \sin \theta_v \hat{\Vec{b}} \Bigl{]}, \label{eq:vthetav} \\
\Vec{r} &=& X \hat{\Vec{a}} + Y \hat{\Vec{b}}, \label{eq:rab}\\
	&=& x \hat{\Vec{v}} + y \hat{\Vec{u}},\label{eq:rvu}\\
\left(
\begin{array}{c}
\hat{\Vec{v}}  \\
\hat{\Vec{u}}
\end{array}
\right)
&\equiv&
\left(
\begin{array}{cc}
\cos \theta_v  & \sin \theta_v \\
- \sin \theta_v & \cos \theta_v
\end{array}
\right)
\left(
\begin{array}{c}
\hat{\Vec{a}}  \\
\hat{\Vec{b}}
\end{array}
\right).
\label{eq:henka}
\end{eqnarray}
Here, $\theta_v$ is the angle between the $a$ axis and the Fermi velocity (See, Fig.~\ref{fig:za}) 
and $m_x$ and $m_y$ are the anisotropic effective masses.
$k_a$ and $k_b$ are the axes fixed to the crystal axes in momentum space.
In this system, 
a point $(k_x,k_y)$ on the Fermi surface with the energy 
$\varepsilon = k_x^2 /(2 m_x) + k_y^2/(2 m_y)$ being the Fermi energy 
is parameterized with $k_x = k \cos \theta$ and $k_y = k \sqrt{m_y/m_x} \sin \theta$.
%%%%%%%%%%%%%%%%%%%%%%
%%%%%%%%%%%%%%%%%%%%%%5
\subsubsection{Isotropic Fermi surface\cite{thesis}}
%%%%%%%%%%%%%%%%%%%%%
First, we assume superconductors with the circular Fermi surface ($m_x=m_y=m$).
Those with the anisotropic Fermi surface will be discussed briefly in \S 3.1.3.

In the case of $\hDelta \hDelta^{\dagger} \propto \hat{\sigma}_0$, we can solve the matrix Riccati equations easily.
Equations (\ref{eq:a}) and (\ref{eq:b}) are to be solved, under the initial conditions:
%%%%%%%%%%%%%%
\begin{eqnarray}
\ha   = \frac{ \hDelta}
	{\omega_n + \sqrt{\omega_n^2 + \frac{1}{2} {\rm tr} \hDelta \hDelta^{\dagger}}
        }, \: \:  \: {\rm for} \: x = -x_{\rm c},
        \label{eq:ia}\\
\hb   = \frac{ \hDelta^{\dagger}}
	{\omega_n + \sqrt{\omega_n^2 + \frac{1}{2} {\rm tr} \hDelta \hDelta^{\dagger}}
        }, \: \:  \:{\rm for} \: x = x_{\rm c}, \: \: \: \:
        \label{eq:ib}
\end{eqnarray}
%%%%%%%%%%%%%%%
where $x_{\rm c}$ is a positive real cut-off. \cite{Kato}

At the vortex core and low energy, the Green function diverges \cite{Kramer}, which owes to the Andreev bound state.
Therefore, we expand the denominator of the Green function around a vortex core.
We call this Kramer-Pesch approximation, 
which is appropriate around a vortex core in a low energy region($| \omega_n| \ll |\Delta_{\infty}|$).
Here, $\Delta_{\infty}$ is a pair-potential in the bulk region.
We consider the perturbation with respect to an impact parameter $y$ and an energy $\omega_n$.
We consider a pair-potential having the form of 
%%%%%%%%%%%%%%%%%%%%
\begin{eqnarray}
\hat{\Delta}(\rv,\kv) 
 &=& f(r) \exp({\rm i} \theta_r) \hat{\Delta}(\kv), \\
 &=& f(r) \exp({\rm i} \theta_v) \frac{x+{\rm i} y}{\sqrt{x^2+y^2}} \hat{\Delta}(\kv),\\
 \hat{\Delta}^{\dagger} (\rv,\kv) 
 &=& f(r)\exp(-{\rm i} \theta_r) \hat{\Delta}^{\dagger}(\kv), \\
 &=& f(r) \exp(-{\rm i} \theta_v) \frac{x-{\rm i} y}{\sqrt{x^2+y^2}} \hat{\Delta}^{\dagger}(\kv). \: \: \: \: \: \:
 \end{eqnarray}
 %%%%%%%%%%%%%%%%%%%%%%%%%
Here, $f(r)=f(\sqrt{x^2+y^2})$ is the spatial variation of pair-potential and $f(0) = 0$, 
$\lim_{r \rightarrow \infty} f(r) = \Delta_{\infty}$ and $\theta_r$ is an angle from the $a$ axis (See, Fig.~\ref{fig:za}).
We expand a pair-potential written as 
%%%%%%%%%%%%%%%%%%%%%%%%%%
\begin{eqnarray}
\hat{\Delta}(\rv,\kv) 
 \sim f(|x|) \exp({\rm i} \theta_v) \: {\rm sign}(x) \left( 1 +{\rm i} \frac{y}{x} \right) \hat{\Delta}(\kv), \label{eq:del1}\\
 = \hat{\Delta}_0 + \hat{\Delta}_1 , 
 \qquad \qquad \qquad \qquad \qquad  \quad  \; \;\\
\hat{\Delta}^{\dagger}(\rv,\kv) 
 \sim f(|x|) \exp(-{\rm i} \theta_v)  \: {\rm sign}(x) \left( 1 - {\rm i} \frac{y}{x} \right)
\hat{\Delta}^{\dagger}(\kv), \label{eq:del2}\\
= \hat{\Delta}^{\dagger}_0 + \hat{\Delta}^{\dagger}_1, 
\qquad \qquad \qquad \qquad \qquad \qquad \quad 
\end{eqnarray}
with respect to $y$.
%%%%%%%%%%%%%%%%%%%
Here, $\hat{\Delta}_1$ is defined as ${\rm i} (y/x) \hat{\Delta}_0$.
We should notice that $\hat{\Delta}_1$ does not have the singularity at $x = 0$ 
since $f(|x|)/x$ is finite everywhere.
Further remarks will be presented in \S 5.

In the zero-th order with respect to $y$ and $\omega_n$, equations (\ref{eq:a}) and (\ref{eq:b}) are written as
\begin{eqnarray}
 v_{\rm F} \frac{\partial \ha_0}{\partial x}   + \ha_0 \hat{\Delta}^{\dagger}_0 \ha_0 - \hat{\Delta}_0 &=& 0,\label{eq:rica0a}\\
 v_{\rm F} \frac{\partial \hb_0}{\partial x}   - \hb_0 \hat{\Delta}_0 \hb_0 + \hat{\Delta}^{\dagger}_0 &=& 0.\label{eq:rica0b}
\end{eqnarray}
Under the initial conditions (\ref{eq:ia}) and (\ref{eq:ib}), we can obtain the solutions:
\begin{eqnarray}
\ha_0 = \frac{\hDelta_0}
	{\sqrt{\frac{1}{2} {\rm tr} \hDelta_0 \hDelta^{\dagger}_0}
        } \Biggl{|}_{x \rightarrow -x_{\rm c}} 
        = \frac{- \hDelta(\kv)\exp({\rm i} \theta_v)}{\lambda(\kv)}, \label{eq:ai}\\
\hb_0 = \frac{ \hDelta^{\dagger}_0}
	{\sqrt{\frac{1}{2} {\rm tr} \hDelta_0 \hDelta^{\dagger}_0}
        }\Biggl{|}_{x \rightarrow x_{\rm c}}
         = \frac{ \hDelta^{\dagger}(\kv)\exp(-{\rm i} \theta_v)}{\lambda(\kv)}, \label{eq:bi}
\end{eqnarray}
where $\lambda(\kv) = \sqrt{\frac{1}{2}{\rm tr} \hDelta(\kv) \hDelta^{\dagger}(\kv)}$.
For the moment, we denote $\lambda(\kv)$ by $\lambda$ for simplicity.

In the first order with respect to $y$ and $\omega_n$, equations (\ref{eq:a}) and (\ref{eq:b}) are written as
%%%%%%%%%%%%%%%%%
\begin{eqnarray}
 v_{\rm F} \frac{\partial \ha_1}{\partial x}  + 2  \omega_n \ha_0 + \ha_0 \hat{\Delta}^{\dagger}_0 \ha_1 +
 \ha_1 \hat{\Delta}^{\dagger}_0 \ha_0 & & \nonumber \\
 + \ha_0 \hat{\Delta}^{\dagger}_1 \ha_0 
  - \hat{\Delta}_1 
  &=& 0,\\
 v_{\rm F} \frac{\partial \hb_1}{\partial x}  - 2  \omega_n \hb_0 - \hb_0 \hat{\Delta}_0 \hb_1 
 - \hb_1 \hat{\Delta}_0 \hb_0 & & \nonumber \\
 -\hb_0 \hat{\Delta}_1 \hb_0 
 + \hat{\Delta}^{\dagger}_1 
 &=& 0.
\end{eqnarray}
%%%%%%%%%%%%%%%%%%
These equations are inhomogeneous linear differential equations.
Therefore, the solutions
%%%%%%%%%%%%%%%%
\begin{eqnarray}
\ha_1 &=&  \frac{2 }{ v_{\rm F}} \exp \left[ 2 \lambda F(x) \right]  \nonumber \\
& & \times \int_{-x_{\rm c}}^{x} dx' \left[ {\rm i} \frac{y}{x'}\hDelta_0  -  \omega_n \ha_0
 \right] 
\exp \left[- 2 \lambda  F(x') \right]  \nonumber \\
& & + \exp \left[ 2 \lambda  \right\{ F(x)-F(-x_{\rm c}) \left\} \right] \exp({\rm i} \theta_v) \nonumber \\
& & \times \frac{(-1) }{\lambda} \left( -{\rm i} \frac{y}{x_{\rm c}} - \frac{\omega_n}{\Delta_{\infty} \lambda} \right) \hDelta(\kv), \\
\hb_1 &=&  \frac{2 }{ v_{\rm F}} \exp \left[ 2 \lambda  F(x) \right] \nonumber \\
& & \times \int_{x_{\rm c}}^{x} dx' \left[ {\rm i} \frac{y}{x'}\hDelta_0^{\dagger}  + \omega_n \hb_0
 \right] 
\exp \left[- 2 \lambda  F(x') \right] \nonumber \\
& & + \exp \left[ 2 \lambda  \right\{ F(x)-F(x_{\rm c}) \left\} \right] \exp(-{\rm i} \theta_v)\nonumber \\
& & \times \frac{1 }{\lambda} \left( -{\rm i} \frac{y}{x_{\rm c}} - \frac{\omega_n}{\Delta_{\infty} \lambda} \right) \hDelta^{\dagger}(\kv)
\end{eqnarray}
%%%%%%%%%%%%%%%%
are obtained by the method of variation of constants.
Here we introduce 
%%%%%%%%%%%%%%55
\begin{eqnarray}
F(x) &=& \frac{1}{v_{\rm F}} \int_0^{|x|} dx' f(x').
\end{eqnarray}
%%%%%%%%%%%%555.
Substituting $\ha   = \ha_0 + \ha_1$ and $\hb   = \hb_0 + \hb_1$ into eq.~(\ref{eq:green}), 
we obtain Green function up to the first order:
\begin{eqnarray}
\hat{g} &\sim& -2(\ha_0 \hb_1 + \ha_1 \hb_0)^{-1}, \\
&=& -\frac{v_{\rm F} e^{- 2 \lambda F(x)}\hat{\sigma}_0}
	{  \int_{- \infty}^{\infty} dx' \left[ {\rm i} \frac{y}{|x'|} \lambda f(|x'|) +  \omega_n
 \right] 
e^{ -2 \lambda F(x') } } , \: \: \: \: \: \: \: \: \\
&=& -\frac{v_{\rm F} e^{- 2 \lambda F(x)}}{ 2 C  \left[ {\rm i} E +  \omega_n
 \right] }\hat{\sigma}_0,\label{eq:g1}
\end{eqnarray}
in the limit $x_{\rm c} \rightarrow \infty$.
 Here 
 \begin{eqnarray}
C(\theta) &=& \int_{0}^{\infty} dx' e^{ -2 \lambda F(x') },\\
E(\theta) &=&\frac{y}{C} \int_0^{\infty} dx'\frac{\lambda f(x')}{x'} e^{ -2 \lambda F(x') },
\end{eqnarray}
and $E$ is independent of $x$.

\begin{figure}[htbp]
 \begin{center}
  \includegraphics[width=6cm,keepaspectratio]{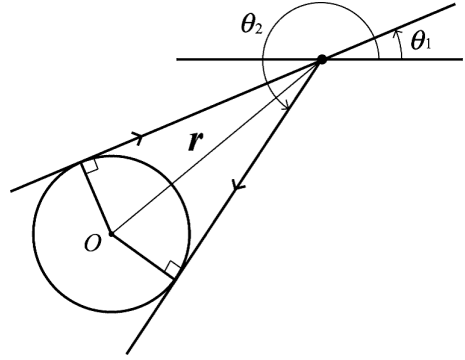}
\caption{Paths that contribute to the integrand of the LDOS at a certain point $\rv$ in real space 
for an isotropic $s$-wave superconductor.
$O$ denotes the vortex center. 
$\theta_1$ and $\theta_2$ are the angles characterizing these paths.}
\label{fig:2path}
 \end{center}
\end{figure}

The retarded Green function $\hat{g}^{\rm R}$ is obtained by replacing ${\rm i} \omega_n$ with $\epsilon + i \delta$ in eq.~(\ref{eq:g1}).
When the Fermi surface is isotropic, $\theta_v = \theta$ (See eqs.~(\ref{eq:vtheta}) and (\ref{eq:vthetav})).
Near the vortex core ($|r| \ll \xi$), we use the following approximations:
\begin{eqnarray}
 \nu (\rv ,\epsilon) &\sim& \lim_{\delta \rightarrow 0} \frac{\nu(0) }{2 } \int_0^{2 \pi}  \frac{ v_{\rm F}d \theta}{2 \pi C} \frac{\delta e^{-2 \lambda F(x)}}{(\epsilon -E)^2 + \delta^2},
 \label{eq:inte}  \\
&\sim& \frac{\nu(0) v_{\rm F}}{  \xi_0} \int_0^{2 \pi}   d \theta  
\lambda \delta (\epsilon - E) e^{-2 \lambda F(x)},\label{eq:deltanu} \: \: \: \: \: \: \: 
\end{eqnarray}
where
\begin{eqnarray}
C \sim \int_0^{\infty} dx' e^{-2 \lambda \frac{\Delta_{\infty} x'}{v_{\rm F}}} &\sim& \frac{\xi_0}{\lambda}, \\
E \sim \frac{y \lambda}{\xi_0} \int_0^{\xi_0} \frac{\Delta_{\infty} \lambda}{\xi_0} dx' &\sim&  \frac{y}{\xi_0} \Delta_{\infty} \lambda^2,
\end{eqnarray}
with $\xi_0 = v_{\rm F}/(\pi \Delta_{\infty})$.
Here, we note that the integrand of $\nu (\rv ,\epsilon)$ in eq.~(\ref{eq:deltanu}) is not 
zero on the path where the relation ($\epsilon - E(\rv,\theta) = 0$) is satisfied.
The number of the solutions $\theta_i$ satisfying $\epsilon - E(\rv,\theta_i) = 0$ is finite as shown in Fig.~\ref{fig:2path}.
We call this path the quasiparticle path.\cite{Ueno}
%%%%%%%%%%%%%%%%%%%%%%%%5

We consider the points at which the LDOS diverges.
We rewrite eq.~(\ref{eq:deltanu}) as 
\begin{equation}
\nu (\rv ,\epsilon) \sim  \frac{\nu(0) v_{\rm F}}{  \xi_0} \int_0^{2 \pi} d \theta \sum_{i} \lambda(\theta)
	\frac{\delta(\theta - \theta_i)}{\Bigl{|} \frac{\partial h(\rv,\theta,\epsilon)}
        {\partial \theta}\Bigl{|}} e^{-2 \lambda(\theta) F(x(\theta))}, 
\end{equation}
\begin{equation}
        = \frac{\nu(0) v_{\rm F}}{ \xi_0} \sum_i \lambda(\theta_i) \Bigl{|} \frac{\partial h(\rv,\theta,\epsilon)}
        	{\partial \theta}\Bigl{|}_{\theta = \theta_i}^{-1} \exp \{ -2 \lambda(\theta_i) F(x(\theta_i)) \},\label{eq:nu2d}
\end{equation}
with $h(\rv, \theta,\epsilon) = \epsilon - E(\rv,\theta)$.
%%%%%%%%%%%%%
%%%%%%%%%%%%%
The condition that the LDOS diverges at a given point $\rv$ for an energy $\epsilon$ is given by
\begin{eqnarray}
h(\rv,\theta,\epsilon) &=& 0 \label{eq:h0}, \\
\frac{\partial h(\rv,\theta,\epsilon)}{\partial \theta} &=& 0.
\label{eq:hassan}
\end{eqnarray}
When we regard eq.~(\ref{eq:h0}) as the equations of a family of lines on $x$-$y$ plane 
with one-parameter $\theta$, equations~(\ref{eq:h0}) and (\ref{eq:hassan}) describe 
the enveloping curve of that family of lines ($=$ the quasiparticle paths in the present case).
From eqs.~(\ref{eq:h0}) and (\ref{eq:hassan}), the parametric equation
%%%%%%%%%%%%%%%%%
\begin{eqnarray}
\tilde{x} &=& 
2 \tilde{\epsilon} \frac{ \frac{\partial \lambda(\theta)}{\partial \theta} }
{\lambda^3(\theta)}, \\
\tilde{y} &=& \frac{\tilde{\epsilon}}{\lambda^2(\theta)}
\end{eqnarray}
follows.
%%%%%%%%%%%%%%%%%%%%%%%
Here, we use dimensionless variables as $\tilde{\epsilon}  = \epsilon / \Delta_{\infty}$, 
$\tilde{x} = x/\xi_0$ and $\tilde{y} = y/\xi_0$ and we regard $\lambda(\theta)$ as $\lambda(\kv)$.
From eqs.~(\ref{eq:rab})-(\ref{eq:henka}), we obtain the parametric equations:
\begin{eqnarray}
\left(
\begin{array}{c}
\tilde{X}  \\
\tilde{Y} 
\end{array}
\right)
=
\frac{\tilde{\epsilon}}{\lambda^2(\theta)} \left(
\begin{array}{c}
2 \frac{\frac{\partial }{\partial \theta} \lambda(\theta)}{\lambda(\theta)} \cos \theta - \sin \theta  \\
 2 \frac{\frac{\partial}{\partial \theta}  \lambda(\theta)}{\lambda(\theta)} \sin \theta + \cos \theta
\end{array}
\right),
\end{eqnarray}
for the enveloping curve, on which the LDOS diverges.

\begin{figure}[htbp]
 \begin{center}
  \includegraphics[width=6cm,keepaspectratio]{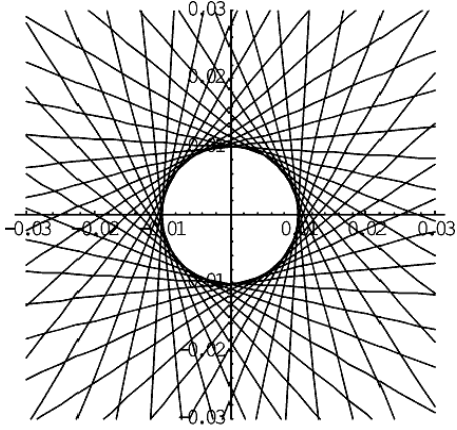}
\caption{Set of quasiparticle paths for $\epsilon/\Delta_{\infty} = 0.01$ in the isotropic $s$-wave superconductor.
 The LDOS diverges on the enveloping curve of the quasiparticle paths, which in the present 
 case forms the circle $\tilde{X}^2+\tilde{Y}^2 = \tilde{\epsilon}^2$.}
\label{fig:tra}
 \end{center}
\end{figure}

%%%%%%%%%5
We consider two examples.
First, we consider isotropic $s$-wave superconductor with circular Fermi surface, where $\lambda(\theta) = 1$ .
In this case, two paths specified by $\theta_1$ and $\theta_2$ contribute to $\nu(\rv,\epsilon)$ in eq.~(\ref{eq:nu2d}), as shown in 
Fig.~\ref{fig:2path}.
The parametric equations for the enveloping curve is given by 
\begin{eqnarray}
\left(
\begin{array}{c}
\tilde{X}  \\
\tilde{Y} 
\end{array}
\right)
=
\tilde{\epsilon} \left(
\begin{array}{c}
 - \sin \theta  \\
 \cos \theta
\end{array}
\right),
\end{eqnarray}
which forms the circle with a radius of $\tilde{\epsilon}$ as shown in Fig.~\ref{fig:tra}.
%%%%%%%%%%%%%%%%%%%55
%%%%%%%%%%%%%%%%%5

\begin{figure}[htbp]
 \begin{center}
  \includegraphics[width=8.5cm,keepaspectratio]{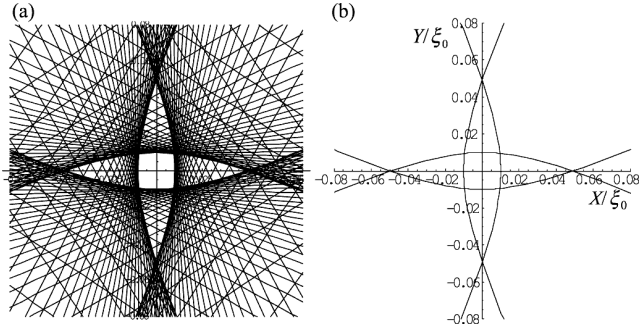}
\caption{Distribution of the LDOS in the $d$-wave superconductor about $\epsilon/\Delta_{\infty} = 0.01$.
 (a): the set of quasiparticle path. (b): the enveloping curve of the quasiparticle paths. 
The LDOS diverges on the enveloping curve of the quasiparticle paths. }
\label{fig:d-wave}
 \end{center}
\end{figure}

Second, we consider the $d$-wave superconductor with the circular Fermi surface; $\lambda(\theta) = \cos 2 \theta$.
In this system, we obtain the following parametric equations:
\begin{eqnarray}
\left(
\begin{array}{c}
\tilde{X}  \\
\tilde{Y} 
\end{array}
\right)
=
\frac{\tilde{\epsilon}}{\cos^2 2 \theta} \left(
\begin{array}{c}
 -4 \cos \theta \tan 2 \theta - \sin \theta  \\
 -4 \sin \theta \tan 2 \theta + \cos \theta
\end{array}
\right),
\end{eqnarray}
where the LDOS diverges.
In Fig.~\ref{fig:d-wave}, we show the set of quasiparticle paths (Fig.~\ref{fig:d-wave}(a)) 
and the enveloping curve of them (Fig.~\ref{fig:d-wave}(b)).
%%%%%%%%%%%%%%%%%%%5
%%%%%%%%%%%%%%%%%%%%%%%
We can see that the enveloping curve 
describes the overall features of the spatial distribution of the LDOS obtained by numerical calculations 
by Schopohl \textit{et al.} \cite{Schopohl2} and Ichioka \textit{et al.} \cite{ichioka}
The LDOS patterns for different energies reduce to a single figure after scaling a spatial coordinate.
Therefore, the LDOS pattern in the limit $\tilde{\epsilon} \rightarrow 0$ is expressed as 
\begin{equation}
\tilde{Y} = \pm \tilde{X}.
\end{equation}
This is the reason why the LDOS pattern rotates by 45 degree with increasing the energy.\cite{ichioka}
%%%%%%%%%%

\begin{figure}[htbp]
 \begin{center}
  \includegraphics[width=8cm,keepaspectratio]{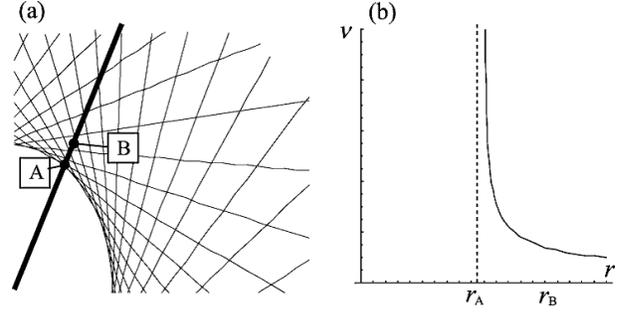}
\caption{(a) Line in the radial direction. (b) Local density of states on the line in the radial direction, 
which has the square-root singularity.}
\label{fig:radial}
 \end{center}
\end{figure}

Now we return to general case and we show that the LDOS has the square-root singularity near the enveloping curves. 
We consider the LDOS on the line running in the radial direction as shown Fig.~\ref{fig:radial}(a).
%%%%%%%%%%%%%%%%%%%%%%
%%%%%%%%%%%%%%%%%%%%%%5
We expand eq.~(\ref{eq:h0}) around a divergence point A$(r_i, \theta_i)$ 
at which the line passes the enveloping curve:
\begin{eqnarray}
h(r,\theta) &\sim& \frac{\partial h}{\partial r} \Bigl{|}_{(r_i,\theta_i)} \Delta r + 
\frac{1}{2} \frac{\partial^2 h}{\partial \theta^2} \Bigl{|}_{(r_i,\theta_i)} (\Delta \theta)^2 \nonumber \\
& & 
+\frac{\partial^2 h}{\partial \theta \partial r} \Bigl{|}_{(r_i,\theta_i)} \Delta r \Delta \theta + 
\frac{1}{2} \frac{\partial^2 h}{\partial r^2} \Bigl{|}_{(r_i,\theta_i)} (\Delta r)^2
. \nonumber \\
\end{eqnarray}
Since $E(r ,\theta) \propto y = r \sin (\theta_r - \theta)$ has the form of $r \times$ (a function of $\theta$), 
it follows that
%%%%%%%%%%%%%%%%
\begin{equation}
\frac{\partial^2 h}{\partial \theta \partial r} \Bigl{|}_{(r_i,\theta_i)} = \frac{1}{r_i} \frac{\partial h}{\partial \theta} \Bigl{|}_{(r_i,\theta_i)} = 0, 
\end{equation}
%%%%%%%%%%%%%%%%%%%%%%
from eq.~(\ref{eq:hassan}) and 
%%%%%%%%%%%%%%%%
\begin{equation}
\frac{\partial^2 h}{\partial r^2} = 0.
\end{equation}
%%%%%%%%%%%%%%%
Hence we obtain
%%%%%%%%%%%%%%%%
\begin{equation}
h(r,\theta) \sim \frac{\partial h}{\partial r} \Bigl{|}_{(r_i,\theta_i)} \Delta r + 
\frac{1}{2} \frac{\partial^2 h}{\partial \theta^2} \Bigl{|}_{(r_i,\theta_i)} (\Delta \theta)^2.\label{eq:hten}
\end{equation}
%%%%%%%%%%%%%%%%%
The solution of eq.~(\ref{eq:h0}) with eq.~(\ref{eq:hten})
%%%%%%%%%%%%%%%%%
\begin{equation}
(\Delta \theta) =\pm \left\{ \left( \left( -2 \frac{\partial h}{\partial r} \Bigl{|}_{(r_i,\theta_i)} \right) \Bigl{/} 
\left( \frac{\partial^2 h}{\partial \theta^2} \Bigl{|}_{(r_i,\theta_i)} \right) \right) \Delta r \right\}^{1/2}
\end{equation}
%%%%%%%%%%%%%%%%%%
exists only when 
%%%%%%%%%%%%%%%5
\begin{equation}
(\Delta r) \frac{\partial h}{\partial r} \Bigl{|}_{(r_i,\theta_i)} 
  \frac{\partial^2 h}{\partial \theta^2} \Bigl{|}_{(r_i,\theta_i)} <0,
\end{equation}
is satisfied.
%%%%%%%%%%%%%
This results from the fact that 
quasiparticle paths exist only in one side of the enveloping curve (See Fig.~\ref{fig:radial}(a)).
We similarly expand eq.~(\ref{eq:hassan}) around $(r_i,\theta_i)$ as
%%%%%%%%%%%%%%%%%%%%
\begin{equation}
\frac{\partial h}{\partial \theta} \sim \frac{\partial^2 h}{\partial \theta^2} 
\Bigl{|}_{(r_i,\theta_i)} \Delta \theta 
+ \frac{\partial^2 h}{\partial \theta \partial r} \Bigl{|}_{(r_i,\theta_i)} \Delta r =\frac{\partial^2 h}{\partial \theta^2} 
\Bigl{|}_{(r_i,\theta_i)} \Delta \theta  .
\label{eq:hbibun}
\end{equation}
%%%%%%%%%%%%%%%%%
From eqs.(\ref{eq:nu2d}), (\ref{eq:hten}) and (\ref{eq:hbibun}), the LDOS is written as
%%%%%%%%%%%%%%%%%%
\begin{equation}
\nu(\rv,\epsilon) \propto \left( - \frac{\partial^2 h}{\partial \theta^2} \Bigl{|}_{(r_i,\theta_i)}  \frac{\partial h}{\partial r} \Bigl{|}_{(r_i,\theta_i)}
	\Delta r  \right)^{-1/2}.
\end{equation}
%%%%%%%%%%%%%%%%%%%
Therefore, the LDOS has the square-root divergence as shown in Fig.~\ref{fig:radial}(b).
%%%%%%%%%%%%%%%%%%%%%%%%%%%%%%55

\subsubsection{Anisotropic Fermi surface}
We consider superconductors with the anisotropic Fermi surface.
Replacing $v_{\rm F}$ with $| \Vec{v}_{\rm F}(\theta)| \equiv v_{\rm F} g(\theta)$ and using eq.~(\ref{eq:ldosa}), 
we can extend the present discussion 
to the anisotropic Fermi surface
because the Eilenberger equation is the differential equation on the line in the direction of the group velocity.\cite{Kopnin}
By replacing $h(\theta)$ with $\epsilon - E(\theta)/g(\theta)$, we see that the points where the LDOS diverges are given by
%%%%%%%%%%%%%%%%%%%
\begin{eqnarray}
\tilde{x} &=& \frac{g \tilde{\epsilon}}{\frac{\partial \theta_v}{\partial \theta} \lambda^2 } \left( 2 \frac{\frac{\partial \lambda}{\partial \theta}}{\lambda}
 - \frac{\frac{\partial g}{\partial \theta}}{g}\right), \\
\tilde{y} &=& \frac{g \tilde{\epsilon}}{\lambda^2}.
\end{eqnarray}
%%%%%%%%%%%%
%%%%%%%%%%%%%%%%%%%%
%%%%%%%%%%%%%%%%%%%5

%%%%%%%%%%%%%%%%%%%%%%%%%%%%%%%%%%%%%%%%%%%%
\subsection{Three-dimensional isotropic Fermi surface}
\begin{figure}[htbp]
 \begin{center}
  \includegraphics[width=8.5cm,keepaspectratio]{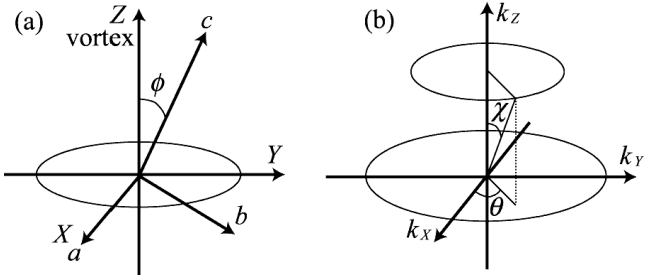}
\caption{Frame of the space. (a): the relation between the coordinates fixed to 
the crystal axis ($a$, $b$ and $c$) and the other coordinates fixed to the magnetic fields ($X$, $Y$ and $Z$).
(b): the planes with different $\chi$.
}
\label{fig:3daxis}
 \end{center}
\end{figure}

%%%%%%%%%%%%%%%%%%%%%%%%%%%%%%%%%%%%%%%%%%%%%
We extend our formulation to superconductors with the three-dimensional Fermi surface.
We consider a single vortex along the $Z$ axis which tilts from the crystal $c$ axis by angle $\phi$ as shown in Fig.~\ref{fig:3daxis}(a).
%%%%%%%%%%%%%%5
%%%%%%%%%%555
%%%%%%%%%%%%%%%%%%%%%%
When we take the $X$ axis on the $a$-$b$ plane, the $Y$ axis is automatically determined. 
We denote by $\hat{\Vec{a}}_{\rm M}$, $\hat{\Vec{b}}_{\rm M}$ and $\hat{\Vec{c}}_{\rm M}$ the unit vectors along $X$, $Y$, $Z$ axis, 
respectively. 
The pair-potential $\hDelta(\rv, \kv)$, $\hDelta^{\dagger}(\rv,\kv)$ in the Riccati equation (\ref{eq:a}) and 
hence $\ha(\rv,\kv)$ and $\hb(\rv,\kv)$ depend on $\rv$ only through $X$ and $Y$ with 
%%%%%%%%%%%%%
\begin{equation}
\rv = X \hat{\Vec{a}}_{\rm M} + Y \hat{\Vec{b}}_{\rm M} + Z \hat{\Vec{c}}_{\rm M},
\end{equation}
%%%%%%%%%%%%
because of a translational symmetry along the $Z$ axis 
(i.e. the direction of vortex). 
As a result, $\Vec{v} \cdot \Vec{\nabla}$ in the Riccati equation can be replaced by 
\begin{equation}
\Vec{v} \cdot \Vec{\nabla} = v_X \frac{\partial}{\partial X} + v_Y \frac{\partial}{\partial Y},
\end{equation}
with $\Vec{v} \cdot \hat{\Vec{a}}_{\rm M} = v_X$ and $\Vec{v} \cdot \hat{\Vec{b}}_{\rm M} = v_Y$. 
Consequently, 
the Riccati equation for three-dimensional system turns into the same form as that for two-dimensional system.
The momentum and the Fermi velocity $\Vec{v}_{\rm F}$ at each point on the Fermi surface are written as 
%%%%%%%%%%%%%%%%%%
\begin{eqnarray}
\tilde{\kv} &=&   \sin \chi \cos \theta  \hat{\Vec{a}}_{\rm M} +  \sin \chi \sin \theta  \hat{\Vec{b}}_{\rm M} \nonumber \\
& & +  \cos \chi \hat{\Vec{c}}_{\rm M},  \\
&=& \tilde{k}_X \hat{\Vec{a}}_{\rm M} + \tilde{k}_Y \hat{\Vec{b}}_{\rm M} + \tilde{k}_Z \hat{\Vec{c}}_{\rm M}, \\
\Vec{v} &=&  v_{\rm F} \hat{\Vec{v}} = v_{\rm F } \Bigl{[} \sin \chi ( \cos \theta \hat{\Vec{a}}_{\rm M} 
 +  \sin \theta \hat{\Vec{b}}_{\rm M})\nonumber \\
& & \; \; \; \; \; \; \; \; \; \; \; +  \cos \chi \hat{\Vec{c}}_{\rm M}  \Bigl{]}.
\end{eqnarray}
As a real-space coordinate on $\Vec{a}_{\rm M}$-$\Vec{b}_{\rm M}$ plane, we introduce $x$, $y$ by 
\begin{equation}
\rv = x \hat{\Vec{v}} + y \hat{\Vec{u}} + Z \hat{\Vec{c}}_{\rm M},
\end{equation}
with 
\begin{equation}
\hat{\Vec{u}} = - \sin \theta \hat{\Vec{a}}_{\rm M} + \cos \theta \hat{\Vec{b}}_{\rm M}.
\end{equation}
The resultant matrix Riccati equations for three-dimensional system are then written as 
%%%%%%%%%%%%%%%%%%%%%%%%%
%%%%%%%%%%%%%%%%%%%%%%%
\begin{eqnarray}
  v_{\rm F } \sin \chi \frac{\partial \ha  }{\partial x} + 2  \omega_n \ha   + \ha   \hat{\Delta}^{\dagger} \ha     -
  \hat{\Delta} &=& 0,
 \\
 v_{\rm F } \sin \chi \frac{\partial \hb}{\partial x} - 2  \omega_n  \hb - \hb \hat{\Delta} \hb + \hat{\Delta}^{\dagger} 
&=& 0,
\end{eqnarray}
with the pair-potential
\begin{eqnarray}
\hat{\Delta}(\rv,\tilde{\kv}) 
 &=& f(r) \exp({\rm i} \theta_r) \hat{\Delta}(\tilde{\kv}), \\
 \hat{\Delta}^{\dagger} (\rv,\tilde{\kv}) 
 &=& f(r)\exp(-{\rm i} \theta_r) \hat{\Delta}^{\dagger}(\tilde{\kv}), \\
 \lambda(\tilde{\kv}) &\equiv& \sqrt{\frac{1}{2} \: {\rm tr} \hDelta(\tilde{\kv}) \hDelta^{\dagger}(\tilde{\kv})}.
 \end{eqnarray}
 Here $r = \sqrt{x^2+y^2}$.
 For the moment, we denote $\lambda(\kv)$ 
 by $\lambda$ for simplicity.
%%%%%%%%%%%%%%%%%%%%%%%%%%%%%%%%%%%%%% 

The $\kv$-dependence of $\hDelta(\tilde{\kv})$ on the Fermi surface 
respects the direction of crystal axes ($a$, $b$, $c$). 
Thus, it is convenient to write $\hDelta(\tilde{\kv}) = \hDelta(\tilde{k}_a,\tilde{k}_b, \tilde{k}_c)$ 
in terms of components of $\tilde{\kv}$ for each crystal axis. 
Since the direction of magnetic field (or vortex) is 
tilted by $\phi$ from the $c$ axis within $b$-$c$ plane, 
$\tilde{k}_a$, $\tilde{k}_b$, $\tilde{k}_c$ are, respectively, written as 
\begin{eqnarray}
\tilde{k}_a &=& \tilde{k}_X = \sin \chi \cos \theta, \\
\tilde{k}_b &=& \tilde{k}_Y \cos \phi - \tilde{k}_Z \sin \phi, \\
           &=& \sin \chi \sin \theta \cos \phi - \cos \chi \sin \phi, \\
\tilde{k}_c &=& \tilde{k}_Y \sin \phi + \tilde{k}_Z \cos \phi, \\
	&=& \sin \chi \sin \theta \sin \phi + \cos \chi \cos \phi. \: \: \: \: \: \: \:
\end{eqnarray}
%%%%%%%%%%%%%%%%%%%%%%%%%%%%%%%%%%%%%%%%%%%%%%%%%%%
Therefore, the pair-potential is rewritten as
%%%%%%%%%%%%%%
\begin{eqnarray}
 \hat{\Delta}(\tilde{\kv})        = \hat{\Delta} \Bigl{(} \sin \chi \cos \theta,
 \sin \chi \sin \theta \cos \phi - \cos \chi \sin \phi,  \nonumber \\ 
 \sin \chi \sin \theta \sin \phi + \cos \chi \cos \phi \Bigl{)}. \label{eq:delta}
\end{eqnarray}
Using eq.~(\ref{eq:delta}), we can obtain the LDOS as a function of $\rv$, $\epsilon$ and $\phi$.
In what follows, $\phi$-dependence of $\hDelta(\tilde{\kv})$ will not be shown explicitly.
%%%%%%%%%%%%%%%%%%%%%%%%%%%%%%%%%%%%%%%%%%%%%%%%%%%%%%%

%%%%%%%%%%%%%%%%%%%%%%%%%%%%%%%%%%%%%%%%
Near the vortex core ($|r| \ll \xi$), the local density of states is given by 
%\begin{eqnarray}
\begin{equation}
\nu (\rv ,\epsilon) =  \lim_{\delta \rightarrow 0} \frac{\nu(0) v_{\rm F}}{ 4 \xi_0} \int_0^{\pi} d \chi \sin \chi \int_0^{2 \pi} 
\frac{d \theta}{2 \pi} \frac{  \lambda \delta e^{-2 \lambda F(x)}}{(\epsilon -E_{\rm 3 D})^2 + \delta^2}, \label{eq:3ddis}
\end{equation}
% \nonumber\\
\begin{equation}
                    \sim \frac{\nu(0) v_{\rm F}}{ 8 \xi_0} \int_0^{\pi} d \chi \sin \chi \int_0^{2 \pi} d \theta \lambda  \delta (\epsilon - E_{\rm 3 D})e^{-2 \lambda F(x)}, \label{eq:3dnud}
\end{equation}
%\end{eqnarray}
with  the approximations
\begin{eqnarray}
C(\theta , \chi)  &\sim& \frac{\xi_0}{|\lambda(\theta, \chi)|} \sin \chi,\label{eq:c3d} \\
E_{\rm 3D}(\theta , \chi) &\sim& \frac{y}{\xi_0 \sin \chi}  \Delta_{\infty} \lambda^2(\theta, \chi).\label{eq:e3d}
\end{eqnarray}
%%%%%%%%%%%%%%%%%%%%%

\begin{figure}[htbp]
 \begin{center}
  \includegraphics[width=8cm,keepaspectratio]{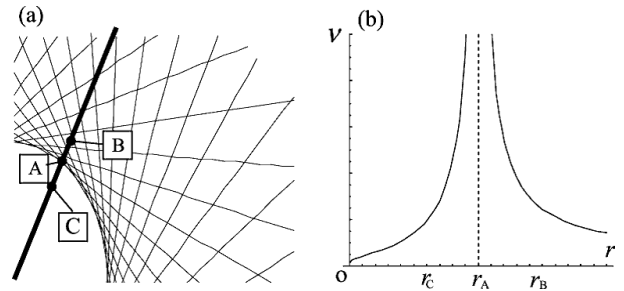}
\caption{(a) Path in the radial direction. The set of the quasiclassical paths for a fixed $\chi$ are drawn. 
(b) Local density of states on the path in the radial direction. 
The singularity is logarithmic.}
\label{fig:3dlog}
 \end{center}
\end{figure}

Now we discuss the condition that $\nu(\rv,\epsilon)$ diverges at a given point $\rv_0$ for an energy $\epsilon$.
Let $(\theta_0,\chi_0,\rv_0)$ be a solution satisfying
\begin{equation}
h(\theta,\chi,\rv) = \epsilon - E_{\rm 3D}(\theta,\chi,\rv) = 0.\label{eq:kato}
\end{equation}
To keep satisfying eq.~(\ref{eq:kato}), $\Delta \theta \equiv \theta - \theta_0$, $\Delta \chi \equiv \chi - \chi_0$ and 
$\Delta \rv \equiv \rv - \rv_0$ have to satisfy
\begin{equation}
\frac{\partial h }{\partial \theta} \Bigl{|}_0 \Delta \theta + \frac{\partial h }{\partial \chi} \Bigl{|}_0 \Delta \chi 
+ \frac{\partial h }{\partial \rv} \Bigl{|}_0 \cdot \Delta \rv = 0, 
\end{equation}
where $\frac{\partial h }{\partial \theta} \Bigl{|}_0 =\frac{\partial h }{\partial \theta} \Bigl{|}_{(\theta_0,\chi_0,\rv_0)}$ and so on.
First we assume that 
\begin{equation}
\left( \frac{\partial h }{\partial \theta} \Bigl{|}_0 \right)^2 
 +\left( \frac{\partial h }{\partial \chi} \Bigl{|}_0 \right)^2 > 0\label{eq:kato2}
\end{equation}
and we introduce a set of new variables
\begin{equation}
\eta \equiv \frac{\partial h }{\partial \theta} \Bigl{|}_0 \Delta \theta + \frac{\partial h }{\partial \chi} \Bigl{|}_0 \Delta \chi , \: \:
\zeta \equiv - \frac{\partial h }{\partial \chi} \Bigl{|}_0 \Delta \theta + \frac{\partial h }{\partial \theta} \Bigl{|}_0 \Delta \chi.
\end{equation}
We then obtain
\begin{equation}
d \theta d \chi \delta(h(\theta,\chi,\rv)) = \frac{\delta \left( \eta + \frac{\partial h }{\partial \rv} \Bigl{|}_0 \cdot \Delta \rv \right) d \eta d \zeta}
{\left( \frac{\partial h }{\partial \theta} \Bigl{|}_0 \right)^2 
 +\left( \frac{\partial h }{\partial \chi} \Bigl{|}_0 \right)^2}.
\end{equation}
The denominator in the right hand side is nonzero under the condition eq.~(\ref{eq:kato2}), 
and therefore $\nu(\rv, \epsilon)$ is finite.
When
\begin{equation}
\frac{\partial h }{\partial \theta} \Bigl{|}_0 = \frac{\partial h }{\partial \chi} \Bigl{|}_0 = 0 \label{eq:l3}
\end{equation}
hold, $\Delta \theta$, $\Delta \chi$ and $\Delta \rv$ have to satisfy
\begin{equation}
a (\Delta \theta)^2 
 + b (\Delta \chi)^2
+ c \Delta \theta \Delta \chi
+ \frac{\partial h}{\partial \rv} \cdot \Delta \rv = 0 \label{eq:kato3},
\end{equation}
with
\begin{equation}
a =\frac{1}{2} \frac{\partial^2 h}{\partial \theta^2} \Bigl{|}_{0}, \: \: \:b =\frac{1}{2}\frac{\partial^2 h}{\partial \chi^2} \Bigl{|}_{0},
\:\:\: c = \frac{\partial^2 h}{\partial \theta \partial \chi} \Bigl{|}_{0},
\end{equation}
in order to satisfy eq.~(\ref{eq:kato}).
In terms of a set of new variables, 
%%%%%%%%%%%
\begin{eqnarray}
\left(
\begin{array}{c}
\Delta \theta  \\
\Delta \chi 
\end{array}
\right)
 &=& \left(
\begin{array}{cc}
\cos \varphi & -\sin \varphi \\
\sin \varphi & \cos \varphi
\end{array}
\right)
\left(
\begin{array}{c}
P  \\ 
Q  
\end{array}
\right), \: \: \: \: \:\:\:\:\:\:\:  \\
\tan 2 \varphi &=& \frac{c}{a-b},
\end{eqnarray}
%%%%%%%%%%%%%
we obtain
%%%%%%%%%%%%%
\begin{eqnarray}
\tilde{\nu}(\rv,\epsilon) &\equiv& \int d \theta d \chi \delta(h(\theta,\chi, \rv)) \\
&=& \int_{V \ni (0,0)} \delta \left( \alpha P^2 + \beta Q^2 + 
	\frac{\partial h}{\partial \rv} \Bigl{|}_0 \cdot \Delta \rv \right) dP dQ, \nonumber \\
\end{eqnarray}
%%%%%%%%%%%%%
with
\begin{eqnarray}
\alpha &=& \frac{1}{2} \left( a + b + \sqrt{(a-b)^2+c^2} \right), \\
\beta &=& \frac{1}{2} \left( a + b - \sqrt{(a-b)^2+c^2} \right).
\end{eqnarray}
$\nu(\rv , \epsilon)$ and $\tilde{\nu}(\rv, \epsilon)$ have the same singularity around $(\theta_0,\chi_0,\rv_0)$. 
Therefore we consider $\tilde{\nu}(\rv,\epsilon)$, instead of $\nu(\rv,\epsilon)$.
First we show that $\tilde{\nu}(\rv=\rv_0,\epsilon)$ does not diverge when $\alpha \beta > 0$; 
when $\alpha > 0$ and $\beta > 0$, $\tilde{\nu}(\rv_0,\epsilon)$ is written as 
%%%%%%%%
\begin{eqnarray}
\tilde{\nu}(\rv_0,\epsilon) &\propto& \int_0^{R_{\rm c}} \int_0^{2 \pi} \delta (R^2) R dR d \Phi, \\
&=& \pi \int_0^{R_{\rm c}^2} \delta (R^2) d(R^2) = \frac{\pi}{2},
\end{eqnarray}
%%%%%%%%%%%%%%
with $\sqrt{\alpha} P = R \cos \Phi$ and $\sqrt{\beta} Q = R \sin \Phi$.
When $\alpha \beta <0$, on the other hand, $\tilde{\nu}(\rv = \rv_0 ,\epsilon)$ diverges logarithmically.
When $\alpha >0$ and $\beta < 0$, $\tilde{\nu}(\rv, \epsilon)$ becomes 
%%%%%%%%%%%%
\begin{eqnarray}
\tilde{\nu}(\rv ,\epsilon) &\sim& \int_0^{R_{\rm c}} \int_0^{2 \pi}
 \frac{R d R d \Phi}{\sqrt{\alpha \beta}} \nonumber \\
 & & \times \delta \left( R^2 \cos 2 \Phi + \frac{\partial h}{\partial \rv} \Bigl{|}_0 \cdot \Delta \rv \right), \\
 &=& \frac{1}{2 \sqrt{ab - \frac{c^2}{4}}} \int_0^{2 \pi}  \frac{d \Phi}{|\cos 2 \Phi|} \nonumber \\
 & &  \times 
 \Theta \left(R_{\rm c}^2 \cos 2 \Phi - \left| \frac{\partial h}{\partial \rv} \Bigl{|}_0 \cdot \Delta \rv \right| \right), \: \: \:
\\
 &\propto& \frac{\ln \left| \frac{\partial h}{\partial \rv} \Bigl{|}_0 \cdot \Delta \rv \right|}{\sqrt{ab - \frac{c^2}{4}}},
\end{eqnarray} 
%%%%%%%%%%%%%%
where $\Theta$ is Heaviside step function.
Note that the condition $\alpha \beta = ab - c^2/4 < 0$ is equivalent to the saddle point condition
\begin{equation}
\frac{\partial^2 h }{\partial \theta^2} \frac{\partial^2 h }{\partial \chi^2} -
  \left( \frac{\partial^2 h}{\partial \theta \partial \chi}\right)^2 < 0, \label{eq:anten}
\end{equation}
for a surface given by $h(\theta,\chi) = const.$
The logarithmic singularity of the LDOS near the enveloping curve is schematically shown in Fig.~\ref{fig:3dlog}
%%%%%%%%%%%%%5
%%%%%%%%%%%%%%%%%
%%%%%%%%%%%%%%%%%%%%5
%%%%%%%%%%%5555

\begin{figure}[htbp]
 \begin{center}
  \includegraphics[width=7cm,keepaspectratio]{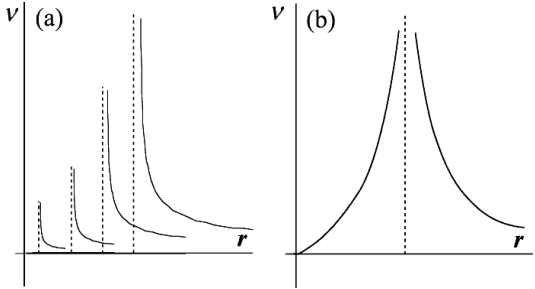}
\caption{(a): Superposition of the divergence point of the LDOS with various $\chi$. 
(b): The LDOS in the three-dimensional system.}
\label{fig:2dto3d}
 \end{center}
\end{figure}

%%%%%%%%%%%%%%%%%%%%%%%%%%%%%%%%%%%%%%%%%%%%%%%
We note that the singularity of the LDOS in three-dimensional system is weaker than that in two-dimensional system. 
Further, in three-dimensional system, the singularity of the LDOS exists in both sides of the 
enveloping curves as seen in Fig.~\ref{fig:3dlog}(b). 
This fact is in contrast to two-dimensional system. 
These two features are understood in the following way. 
The LDOS in three-dimensional system can be regarded as the 
sum of the LDOS of fictitious two-dimensional system perpendicular to the $Z$ axis with reduced Fermi velocity $v_{\rm F} \sin \chi$. 
Repeating the discussion on two-dimensional system, at the points where 
\begin{eqnarray}
\tilde{x} &=& 2 \frac{\tilde{\epsilon} \sin \chi \frac{\partial \lambda}{\partial \theta}}{\lambda^3}, \label{eq:xdiv}\\
\tilde{y} &=& \frac{ \tilde{\epsilon} \sin \chi}{\lambda^2}
\label{eq:ydiv}
\end{eqnarray}
are satisfied, a singular contribution comes from each two-dimensional system. 
Each two-dimensional system has a different Fermi velocity $v_{\rm F} \sin \chi$. 
Hence, singular contributions are gradually shifted in real space (Fig.~\ref{fig:2dto3d}(a)). 
Owing to the integration with respect to $\chi$, the 
resultant LDOS becomes less singular but the singular behavior exists in both sides of an enveloping curve (Fig.~\ref{fig:2dto3d}(b)). 
%%%%%%%%%%%%%%%%%%%%%%%%%%%%%%%%%%%%%%%%%%%%%%%
%%%%%%%%%%%%%%%%%%%%%%%%%%%%%%%%%%%%%%%%%%%%
%%%%%%%%%%%%%%%%%%%%%%%%%%%%%%%%%%%%%%%%%%%

\begin{figure}[htbp]
 \begin{center}
  \includegraphics[width=6cm,keepaspectratio]{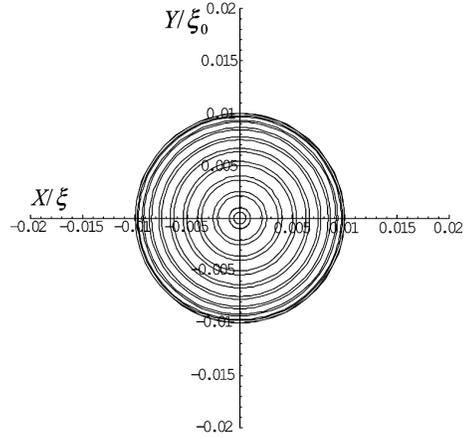}
\caption{Superposition of the LDOS patterns with various $\chi$.}
\label{fig:kasane}
 \end{center}
\end{figure}

For example, we consider an isotropic $s$-wave superconductor with spherical Fermi surface ($\lambda(\theta, \chi) = 1$). 
In this system, we obtain the parametric equations for a fixed $\chi$:
%%%%%%%%%%%%%%%%%%%%%%%%
\begin{eqnarray}
\left(
\begin{array}{c}
\tilde{X}  \\
\tilde{Y} 
\end{array}
\right)
=
\tilde{\epsilon} \sin \chi \left(
\begin{array}{c}
 - \sin \theta  \\
 \cos \theta
\end{array}
\right),
\label{eq:two}
\end{eqnarray}
where the LDOS would diverge if the system were two-dimensional.
In Fig.~\ref{fig:kasane}, we show the superposition of the patterns with various $\chi$ (Fig.~\ref{fig:3daxis}(b)).
%%%%%%%%%%%%%%%%%%%%%%%%%%%%%%%%%%%%%%%%%%%%
%%%%%%%%%%%%%%%%%%%%%%%%%%%%%%%%%%%%%%%%%%%
From the LDOS patterns with various $\chi$ planes, we can infer the points where the LDOS diverges in the three-dimensional system.
From Fig.~\ref{fig:kasane}, we see that these points are on the circle with a radius of $\tilde{\epsilon}$ in this system.
%%%%%%%%%%%%%%%framing%%%%%%%%%%
%%%%%%%%%%%%%%%%%%%%%%%%%%%%%%%%%%%%%%%%%%%%%%5
%%%%%%%%%%%%%%%%%%%%%%%%%%%%%%%%%%%%%%%%%%%%%%%

In the case of the two-dimensional LDOS pattern on the plane fixed to $\chi$, 
we solve 
\begin{equation}
h(\theta,\chi) = \tilde{\epsilon} - \frac{\tilde{y}}{\sin \chi} \lambda^2(\theta,\chi) 
	= 0, 
        \end{equation}
        \begin{equation}
\frac{\partial h}{\partial \theta} = \frac{\tilde{x}}{\sin \chi} \lambda^2(\theta,\chi) - 
\frac{2 \tilde{y} \frac{\partial \lambda(\theta,\chi)}{\partial \theta}}{\sin \chi}   \lambda(\theta,\chi) 
	= 0.
\end{equation}
The equations $\partial h / \partial \chi = 0$ and eq.~(\ref{eq:anten}) 
can be regarded as a consequence of three-dimensionality of the Fermi surface. 
These equations relate $\theta$ and $\chi$. 
For example, in the system where the pair-potential does not depend on $\chi$, 
$\partial h / \partial \chi = 0$ is written as
%%%%%%%%%%%%%%%%%%%%%%
\begin{equation}
\frac{\partial h}{\partial \chi} = \tilde{\epsilon} \frac{ \cos \chi }{\sin \chi} = 0,
\end{equation}
%%%%%%%%%%%%%%%%55
with $\tilde{y} = \tilde{\epsilon} \sin \chi / \lambda^2$.
Therefore, we can understand that $\chi= \pi / 2$ satisfies this relation.
The saddle point condition (\ref{eq:anten}) is written as
%%%%%%%%%%%%%%%%%%%%
\begin{eqnarray}
\frac{\partial^2 h}{\partial \theta^2} \frac{\partial^2 h}{\partial \chi^2} -  \left( \frac{\partial^2 h}{\partial \theta \partial \chi } 
\right)^2 = - \tilde{\epsilon}^2 <0,
\end{eqnarray}
%%%%%%%%%%%%%%%%
with $\chi = \pi / 2$ and $\tilde{y} = \tilde{\epsilon} \sin \chi / \lambda^2$, which is always satisfied.
Therefore, if the pair-potential does not depend on $\chi$, 
the points where the LDOS with three-dimensional Fermi surface diverges are equal to 
the points where the two-dimensional LDOS on the $\chi = \pi / 2$ plane diverges.
In the isotropic $s$-wave superconductor with spherical Fermi surface, 
the LDOS pattern is the circle with radius of $\tilde{\epsilon}$ from eq.~(\ref{eq:two}).
If the pair-potential depends on $\chi$, the LDOS patten is more complicated.

When we consider the anisotropic Fermi surface in the three-dimensional system, 
we just have to replace $v_{\rm F}$ with $|\Vec{v}_{\rm F}(\theta)| \equiv v_{\rm F} g(\theta, \chi)$.
As in the case of two-dimensional system, 
we can apply our formulation 
to the case of anisotropic Fermi surface.

%%%%%%%%%%%%%%%%%%%%%%%%%%%%%%%%%%%%%%%%%%%%%%%%%%%%%
\section{Local Density of States}
%%%%%%%%%%%%%%%%%%%%%%%%%%%%%%%%
%%%%%%%%%%%%%%%%%%%%%%
In ref.~\citen{nagai}, we showed the LDOS patterns on the heavy Fermion superconductor without inversion symmetry CePt$_3$Si, 
the calculation of which is outlined in Appendix B.
In the present section, we show the LDOS patterns on the various materials such as NbSe$_2$ and ${\rm YNi_2B_2C}$.
We use eq.~(\ref{eq:inte}) or eq.~(\ref{eq:3ddis}) with finite $\delta$ to calculate the distribution of the LDOS with 
the numerical integration of eq.~(\ref{eq:deltanu}) or eq.~(\ref{eq:3dnud}).

%%%%%%%%%%%%%%%%%%%%%%%%%%%%%%%%%%
\subsection{Case of ${\it NbSe_2}$: singlet Cooper pairing and two-dimensional Fermi surface}
\begin{figure}[htbp]
 \begin{center}
  \includegraphics[width=8.5cm,keepaspectratio]{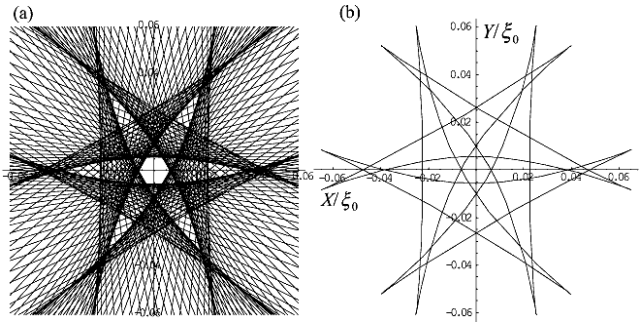}
\caption{Distribution of the LDOS in ${\rm NbSe_2}$ about $\epsilon/\Delta_{\infty} = 0.01$.
 (a): the set of quasiparticle path. (b): the enveloping curve of the quasiparticle paths. 
the LDOS diverges on the enveloping curve of the quasiparticle paths.}
\label{fig:nb}
 \end{center}
\end{figure}

\begin{figure}[htbp]
 \begin{center}
  \includegraphics[width=5cm,keepaspectratio]{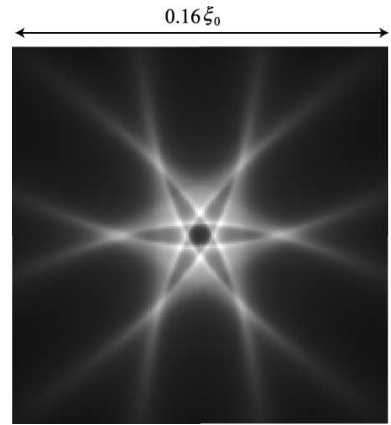}
\caption{Distribtion of the local density of states about ${\rm NbSe_2}$. Here, $\epsilon/\Delta_{\infty} = 0.01$ and 
the smearing factor $\delta/\Delta_{\infty} = 0.001$.
The horizontal axis corresponds to the $a$ axis.}
\label{fig:nbl}
 \end{center}
\end{figure}

We assume that ${\rm NbSe_2}$ has an anisotropic $s$-wave pairing, which is written as
\begin{eqnarray}
\hat{\Delta}(\rv,\kv) &=& i \hat{\sigma}_y \psi(\rv,\kv), \\
\psi(\rv,\kv) &=& |\Delta(\rv)| \exp(i \theta_r) (1 + c_{\rm A} \cos 6 \theta). \: \: \: \: \:
\end{eqnarray}
Here, $\theta_r$ is the phase in real space and $c_{\rm A}$ denotes the degree of anisotropy in the superconducting energy gap.
The experimental data of scanning tunneling spectroscopy indicate that $c_{\rm A} \sim 1/3$.\cite{hayashi}
In the magnetic field parallel to the $c$ axis, we obtain $\lambda$ written as
\begin{equation}
\lambda = 1 + \frac{1}{3} \cos 6 \theta.
\end{equation}
We assume the two-dimensional Fermi surface because ${\rm NbSe_2}$ is a layered superconductor.
We take $g=1$ for simplicity. 
Figure \ref{fig:nb}(a) shows the set of quasiclassical paths of Andreev bound states with energy $\epsilon = 0.01 \Delta_{\infty}$.
The parametric equations for the enveloping curves are given by
\begin{eqnarray}
\tilde{X} = \frac{\tilde{\epsilon}}{(1 + \frac{1}{3} \cos 6 \theta)^2} 
\left(-4 \frac{\cos \theta \sin 6 \theta}{(1 + \frac{1}{3} \cos 6 \theta)}  - \sin \theta \right),\: \: \: \\
\tilde{Y} = \frac{\tilde{\epsilon}}{(1 + \frac{1}{3} \cos 6 \theta)^2} 
\left(-4 \frac{\sin \theta \sin 6 \theta}{(1 + \frac{1}{3} \cos 6 \theta)}  + \cos \theta \right). \: \: \:
\end{eqnarray}
We show the points where the LDOS diverges in Fig.~\ref{fig:nb}(b).
In Fig.~\ref{fig:nbl}, we show the distribution of the LDOS with the numerical integration of eq.~(\ref{eq:inte}).
%%%%%%%%%%%%%%%%%%%%
%%%%%%%%%%%%%%%%%%%
Our LDOS pattern is consistent with earlier numerical results on the LDOS pattern 
calculated by Hayashi \textit{et al.}\cite{hayashi}
%%%%%%%%%%%%%%%%%%%%%%%%%%%%%%%%%%%%%%
\subsection{Case of ${\it YNi_2B_2C}$: singlet Cooper pairing and three-dimensional Fermi surface}
\begin{figure}[htbp]
 \begin{center}
  \includegraphics[width=6cm,keepaspectratio]{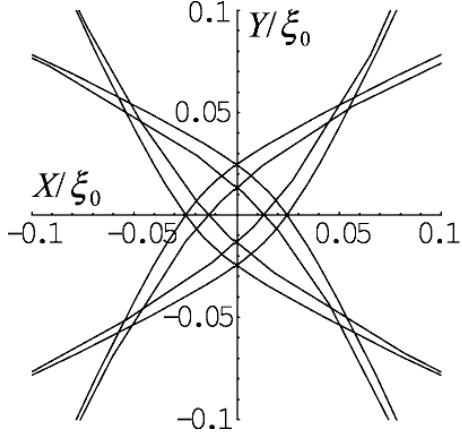}
\caption{Distribution of the points where the local density of states diverges.
The pair-potential is taken to be eqs.~(\ref{eq:yni1})-(\ref{eq:yni3}), 
which is related to ${\rm YNi_2B_2C}$. Here, $\epsilon/\Delta_{\infty} = 0.01$.}
\label{fig:yni}
 \end{center}
\end{figure}

%%%%%%%%%%%%%%%%%%%%%%%%%%%%%%%%%%%5
Next, we consider ${\rm YNi_2B_2C}$.
The LDOS was probed by STM in this material.\cite{Nishimori}
We assume that ${\rm YNi_2B_2C}$ has point nodes in quasiparticle gaps.\cite{Izawa,Maki,Watanabe}
The pair-potential is written as
%%%%%%%%%%%%
\begin{equation}
\hat{\Delta}(\rv,\kv) = i \hat{\sigma}_y \psi(\rv,\kv), \label{eq:yni1}
\end{equation}
\begin{equation}
\psi(\rv,\kv) = |\Delta(\rv)| \exp(i \theta_r) \frac{1}{2} (1 - \sin^4 \chi \cos 4 \theta ), 
\end{equation}
\begin{equation}
\lambda = \frac{1}{2} (1 - \sin^4 \chi \cos 4 \theta ).\label{eq:yni3} 
\end{equation}
%%%%%%%%%%%%
Here, we assume the spherical Fermi surface ($g=1$) and we consider the magnetic field parallel to the $c$ axis.
From eq.~(\ref{eq:l3}), we obtain two parametric equations for enveloping curves 
because the number of the solutions of $\partial h/\partial \chi=0$ is two.
One solution is given by $\cos \chi = 0$ so that the parametric equation is written as
%%%%%%%%%%%%%%
\begin{eqnarray}
\tilde{X} = \frac{4 \tilde{\epsilon}}{(1 -  \cos 4 \theta )^2} 
\left(8 \frac{ \sin 4 \theta \cos \theta}{(1 -  \cos 4 \theta )}  - \sin \theta \right),\label{eq:ynihas1}\\
\tilde{Y} = \frac{4 \tilde{\epsilon}}{(1 -  \cos 4 \theta )^2} 
\left(8 \frac{ \sin 4 \theta \sin \theta}{(1 -  \cos 4 \theta )}  + \cos \theta \right).\label{eq:ynihas2} 
\end{eqnarray}
%%%%%%%%%%%%%5
The other solution is given by $\sin^4 \chi = -(7 \cos 4 \theta)^{-1}$ so that the other parametric equation is written as
%%%%%%%%%%%%%%%
\begin{eqnarray}
\tilde{X} = -\frac{ 49}{16} \tilde{\epsilon} \left( -\frac{1}{7 \cos 4 \theta} \right)^{1/4}\left(
\tan 4 \theta \cos \theta +   \sin \theta \right),\label{eq:ynihas3}\\
\tilde{Y} = -\frac{ 49}{16}\tilde{\epsilon} \left( -\frac{1}{7 \cos 4 \theta} \right)^{1/4} \left(
\tan 4 \theta \sin \theta -   \cos \theta \right),\label{eq:ynihas4}\; 
\end{eqnarray}
%%%%%%%%%%%
with
%%%%%%%%%%%%
\begin{equation}
0 < -\frac{1}{7 \cos 4 \theta} < 1.
\label{eq:cos}
\end{equation}
%%%%%%%%%%5
The points where the LDOS diverges are two curves shown in Fig.~\ref{fig:yni}.
One curve runs away to infinity, while the other curve stays within a finite distance from the origin.
%%%%%%%%%%%%%%%%%%%%%%%%
%%%%%%%%%%%%%%%55
\begin{figure}[htbp]
 \begin{center}
  \includegraphics[width=8cm,keepaspectratio]{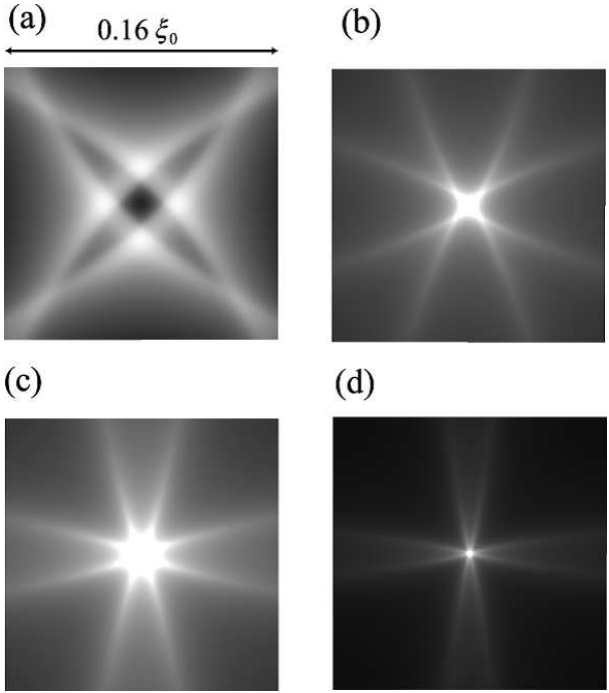}
\caption{Distribtion of the local density of states for the vortex parallel to the $c$ axis about ${\rm YNi_2B_2C}$. 
(a) $\epsilon/\Delta_{\infty} = 1.0 \times 10^{-2}$ and the smearing factor $\delta/\Delta_{\infty}= 0.25 \times 10^{-2}$: 
(b) $\epsilon/\Delta_{\infty} = 1.0 \times 10^{-3}$ and $\delta/\Delta_{\infty} = 0.5 \times 10^{-3}$: 
(c) $\epsilon/\Delta_{\infty} = 1.0 \times 10^{-4}$ and $\delta/\Delta_{\infty} = 0.95 \times 10^{-4}$: 
(d) $\epsilon/\Delta_{\infty} = 1.0 \times 10^{-5}$ and $\delta/\Delta_{\infty} = 0.95 \times 10^{-5}$.
The horizontal and vertical axes correspond to the $a$ and $b$ axes, respectively.}
\label{fig:ynil}
 \end{center}
\end{figure}
%%%%%%%%%%%%%%%%%%%%%%%%%%%%%%%%%%%%%%
We also calculate the distribution of the LDOS as shown in Fig.~\ref{fig:ynil} with eq.~(\ref{eq:3ddis}).
The only lines approaching to infinity are experimentally observed at zero-energy in the case of the superconductors with gap nodes.
In the case of YNi$_2$B$_2$C, for example, the cross-shaped LDOS pattern can be observed at zero-energy, where the nodes exist at 
$(\theta, \chi)= (0,\pi/2)$, $(\pi/2,\pi/2)$, $(\pi , \pi/2)$ and $(3 \pi/2 , \pi/2)$ (See, Fig.~\ref{fig:ynil}(d)).
With energy increasing, the cross-shaped LDOS pattern rotates by 45 degree (See, Fig.~\ref{fig:ynil}(b) and (d)).

\begin{figure}[htbp]
 \begin{center}
  \includegraphics[width=8cm,keepaspectratio]{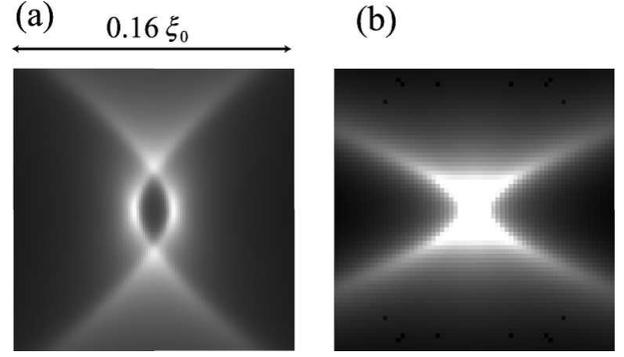}
\caption{Distribtion of the local density of states for the vortex parallel to the $b$ axis about ${\rm YNi_2B_2C}$.
(a) $\epsilon/\Delta_{\infty} = 0.01$ and the smearing factor $\delta/\Delta_{\infty}= 0.003$: 
(b) $\epsilon/\Delta_{\infty} = 5.0 \times 10^{-4}$ and $\delta/\Delta_{\infty} = 5.0 \times 10^{-4}$.
The horizontal and vertical axes correspond to the $a$ and $c$ axes, respectively.}
\label{fig:gyaku}
 \end{center}
\end{figure}

We find that the contribution of the equator of the Fermi surface ($\chi = \pi/2$) to the LDOS pattern is extremely large 
in superconductors with three-dimensional Fermi surface.
Therefore the LDOS pattern of YNi$_2$B$_2$C is similar to that of two-dimensional $d$-wave superconductor.
By the rotation of the magnetic fields, one can obtain the three-dimensional infomation of the pairing symmetry.
For example, in the magnetic field parallel to the $a$ axis, the LDOS pattern of YNi$_2$B$_2$C reflects the two-nodes property of the equator on the Fermi surface as shown in Fig.~\ref{fig:gyaku}.
If YNi$_2$B$_2$C is a $d$-wave superconductor, 
a line node is located along the equator and the LDOS looks different from that shown in Fig.~\ref{fig:gyaku}.

%%%%%%%%%%%%%%%%%%%%%%%%
%%%%%%%%%%%%%%%%%%%%%%%%%%%%%%%%%%%%%%%%%%%%%%%%%%%%%%%%%%%%%%%%%%%%%%%%%%%%%%%%
%%%%%%%%%%%%%%%%%%%%%%%%
\section{Remarks on the Formulation}
We expanded the coefficients of Riccati equations $a(x, y ,\epsilon)$ 
written as 
%%%%%%%%%%%%%%%%%%
\begin{equation}
a(x,y,\epsilon) \sim \frac{\partial a(x,y=0,\epsilon=0)}{\partial y}  y 
+ \frac{\partial a(x,y=0,\epsilon=0)}{\partial \epsilon}  \epsilon,
\end{equation}
%%%%%%%%%%%%%%%%%%
with respect to $y$ and $\epsilon$.
This expansion is valid when $|y|/\xi_0 \ll 1$ and $|\epsilon|/\Delta_{\infty} \ll 1$ are satisfied, 
since typical length and energy are, respectively, $\xi_0$ and $\Delta_{\infty}$.
Although it seems that the expansion of a pair-potential written as eqs.~(\ref{eq:del1}) and (\ref{eq:del2}) with respect to $y$ 
is dangerous near $x = 0$, we note that the dimensionless expansion parameter is not $y/x$ but $y/\xi_0$.
It should be noted that the appropriate region of our approximation is $|x|/\xi_0 \leq 1$
 because the perturbations of $\ha$ and $\hb$ can be applied in the regions $-\infty < x \leq \xi_0$ and $-\xi_0 \leq x < \infty$, respectively.
Practically, this condition does not matter so much 
since the LDOS decreaces exponentially as $\exp(-|r|/\xi_0)$ away from the vortex center.

In our analytical theory, there are two advantages in comparison with direct numerical calculations of Riccati equations.
First, we can obtain physical interpretations of the LDOS pattern. 
The LDOS pattern consists of the enveloping curve of the quasiclassical paths.
We can see from which part of the Fermi surface large LDOS at a point comes from.
Second, we can calculate the LDOS in any anisotropic superconductors 
and in any directions of magnetic fields, 
analytically more easily than numerically.

We calculated the local density of states around a vortex core in the cases of NbSe$_2$ and YNi$_2$B$_2$C.
From eqs.~(\ref{eq:xdiv}) and (\ref{eq:ydiv}), we see that the LDOS patterns are in proportion to $\tilde{\epsilon}$.
The LDOS diverges at the origin at zero-energy.
If the energy $\tilde{\epsilon}$ becomes large, the size of the LDOS patterns becomes large.
Therefore, one can observe more detailed structure of the LDOS pattern with higher energy region.

Besides the LDOS, we can also calculate the physical quantities expressed as an one-particle operator 
because the quasiclassical Green function around a vortex core is derived.
For example, the distribution of the current density around a vortex core is written as
%%%%%%%%%%%%%%%%%5
\begin{equation}
\Vec{j}(\Vec{r}) = 2 e i \pi T \sum_n 
\int \frac{d S_k}{2 \pi^2} \frac{\Vec{v}_{\rm F}}{v_{\rm F}} 
 {\rm Tr} \hat{g}(\hat{\kv},\Vec{r}; i \omega_n).
\end{equation}
%%%%%%%%%%%%%%%%%%

In real materials, the LDOS does not diverge because of the smearing effect.
Both effects of phonons and impurities contribute to the smearing effect.
Therefore, the smearing factor $\delta$ in eq.~(\ref{eq:inte}) or (\ref{eq:3ddis}) determines the cleanness of the LDOS pattern.
The smearing factor can be estimated from the coherence length and the mean free path. 
Since the smearing factor enters into Green function only through $\tilde{\epsilon} + {\rm i} \delta/\Delta_{\infty}$, 
the LDOS depends mainly on $\tilde{\epsilon}$ or $\delta/\Delta_{\infty}$, whichever is larger. 
In the low-energy limit $\tilde{\epsilon} \ll \delta/\Delta_{\infty}$, the LDOS patterns do not depend on $\tilde{\epsilon}$ 
because of the smearing effect; 
zero bias experimental data should be compared with our result in clean limit for energy comparable to 
electron-phonon relaxation rate or impurity scattering rate.

In Appendix B, we show how to solve the Riccati equation in the case of CePt$_3$Si 
in which the Fermi surface is split by spin-degeneracy lifting due to a spin-orbit coupling.
This method has been used in ref.~\citen{nagai}.
However, diagonalizing the Gor'kov equation in the spin space before the application of the quasiclassical theory, 
we obtain the two diagonalized Riccati equations corresponding to the split Fermi surface by spin-orbit coupling.
We do not assume $\alpha \ll \Delta$ in the method diagonalizing the Gor'kov equation 
($\alpha$ represents the strength of the spin-orbit coupling) and 
the resultant Green function in this diagonalizing method is equivalent to the Green function obtained in Appendix B.
Therefore, our theoretical formulation has a wider applicability.
%%%%%%%%%%%%%%%%%%%%%555
%%%%%%%%%%%%%%%%%%%%%%%%%%%%%%%%%%%%%%%%%%%%%%%%%%%%%%%%%%%%%%%%%%%%%%%%%%%%%%%%%%%%%%
\section{Comparison with the STM Experiment}
We compare our results with the experiment with STM by Nishimori {\it et al.}\cite{Nishimori}
They observed the LDOS patterns by STM in the isolated vortex of YNi$_2$B$_2$C and 
found that the LDOS around a vortex core is fourfold-symmetric star-shaped in real space (See, e.g., Fig.~2 in ref.~\citen{Nishimori} ).
The band width $W$ obtained by the numerical calculation of the band structure in YNi$_2$B$_2$C \cite{Ravindran} 
is $W \sim 1$eV and the gap energy is $\Delta \sim 1$meV 
so that $  \Delta/W \sim 1/1000$.
This suggests that the quasiclassical theory is appliable to YNi$_2$B$_2$C.
The existence of gap nodes makes the energy levels continuous, 
according to a quasiclassical calculation supplemented by the Bohr-Sommerfeld-Wilson quantization condition.\cite{Ueno}
Further, the result about the lowest energy levels of the bound state calculated by Hayashi {\it et al.}\cite{hayashi4} 
cannot be applied to YNi$_2$B$_2$C because their calculation is in the case of isotropic $s$-wave superconductor.
Therefore, 
it is not clear whether the STM experiment on YNi$_2$B$_2$C is the observation of quantum regime vortex core.
The size of the core structure in this STM experiment is comparable to the coherence length $\xi_0$.
On the other hand, the appropriate region of our approximation is $r \ll \xi_0$.
It might be beyond the scope of our theory to compare our results with experimental results. 
However, the LDOS pattern by our calculation about NbSe$_2$ is consistent with 
the pattern of the numerical calculation by Hayashi \textit{et al}.\cite{hayashi}, 
which is consistent with 
the STM experiment by Hess {\it et al.}\cite{Hess}
Therefore, we believe that 
it is physically meaningful to compare our results with experimental results.

\begin{figure}[htbp]
 \begin{center}
  \includegraphics[width=4cm,keepaspectratio]{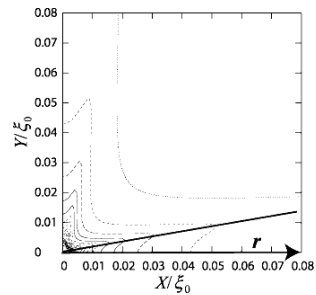}
\caption{Radial dependence of the width of the arm of the star-shape LDOS pattern about YNi$_2$B$_2$C.
$\epsilon/\Delta_{\infty} = 1.0 \times 10^{-5}$ and $\delta/\Delta_{\infty} = 0.95 \times 10^{-5}$.
The angular dependence of the LDOS does not depend on the radial distance from the core.}
\label{fig:angular}
 \end{center}
\end{figure}

\begin{figure}[htbp]
 \begin{center}
  \includegraphics[width=4cm,keepaspectratio]{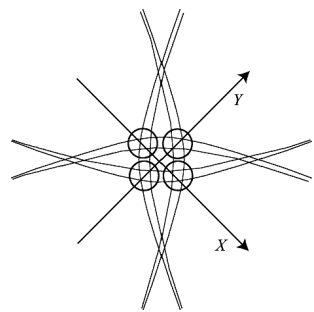}
\caption{LDOS pattern in the large energy region. The intensity of the LDOS is large near the four circles.}
\label{fig:window}
 \end{center}
\end{figure}

From Fig.~3(b) in ref.~\citen{Nishimori}, it is noticed that the angular dependence of $N_s$($0$ meV,$r$,$\theta$) does not 
depend on the radial distance from the core.
This means that the arm of the star-shape pattern spreads.
From eqs.~(\ref{eq:ynihas1})-(\ref{eq:ynihas4}), the arm of the star-shape pattern does not spread in the zero-energy limit.
However, considering the smearing effect, the LDOS pattern in the zero-energy limit with the smearing factor $\delta$ 
is similar to the pattern with the energy $\epsilon = \delta$. 
(Refer to the discussion in \S5.) 
Therefore, we compare the LDOS shown in Fig.~\ref{fig:ynil}(d) with the experimental result.
We can find that the arm of the LDOS pattern shown in Fig.~\ref{fig:ynil}(d) spreads. 
As shown in Fig.~\ref{fig:angular}, the angular dependence of the LDOS does not depend on the radial distance from the core.
Thus, 
the spreading star-shape pattern can be understood by our theory.
In the large energy region, on the other hand, the location of each peak of the LDOS is different from 
that in the STM experiment (See, Fig.~\ref{fig:window} in this paper and Fig.~5(d) in ref.~\citen{Nishimori}).
This suggests that more realistic approximation is necessary for more qualitative comparison with the STM experiment.
For example, we have to calculate the LDOS with the anisotropic Fermi surface, 
for example, tight-binding model.

Our results show that the pairing symmetry itself can be investigated 
by the detailed observation near a single vortex core ($r \ll \xi_0$).
The past experiments by STM have been made in the region $r \sim \xi$.
In this region, it is hard to investigate rich structure of the LDOS pattern because of the exponentially reduction 
of the intensity away from the core.
Therefore, If the LDOS in much narrower energy-range near zero energy and with much higher spatial resolution 
are observed, more detailed information of the pairing symmetry will be obtained.

%%%%%%%%%%%%%%%%%%%%%%%
%%%%%%%%%%%%%%%%%%%%%%
%%%%%%%%%%%%%%%%%%%%%%%%%%%%%%%%%%%%%%%%%%%%%%%
%%%%%%%%%%%%%%%%%%%%%%%55
\section{Conclusions}
We investigated the local density of states (LDOS) around a vortex core on the basis of the quasiclassical theory of superconductivity.
We obtained the analytical formulation of the LDOS in any anisotropic superconductors 
and in any directions of magnetic fields with the use of Kramer-Pesch approximation.
We calculated the LDOS pattern of NbSe$_2$ and YNi$_2$B$_2$C.
The LDOS pattern of NbSe$_2$ is sixfold-symmetric star-shaped and consistent with the numerical calculation by Hayashi \textit{et al.}\cite{hayashi}
The LDOS pattern of YNi$_2$B$_2$C is fourfold-symmetric star-shaped near the zero-energy, 
which is qualitatively consistent with the STM experiment.
Further, without invoking to quantum effect, 
we exhibited the possible reason why the low-energy quasiparticle states spreading 
around the $a$ axis direction were observed in the STM experiment. 
The Band width $W$ estimated from a band calculation \cite{Ravindran}
and the experimentally observed gap energy show that $\Delta/W \sim 1/1000$, and 
therefore YNi$_2$B$_2$C seems not to be in the quantum regime. 
On the other hand, if the LDOS pattern of YNi$_2$B$_2$C in the magnetic field parallel to the $a$ axis is similar to Fig.~\ref{fig:gyaku}, 
YNi$_2$B$_2$C is likely to be a point-node superconductor.
Our results explicitly demonstrate the way to determine the anisotropy of pairing-potential in superconductors from STM experiments.

\section*{Acknowledgment}
This work is supported by Grant-in-Aid for Scientific Research (C)(2)
No.17540314 from JSPS.
%%%%%%%%%%%%%%
%%%%%%%%%%%%%%
%%%%%%%%%%%%%%%%%%%%%%%%%%%%%%%%%%%%%%%%%%%
\appendix
\section{Projection Method}
In this Appendix, we simplify the Eilenberger equation with a parametrization for the propagators, 
which were originally introduced in the study of the vortex dynamics by Eschrig. \cite{eschrig,eschrig2}
First, we define the projectors written as
\begin{equation}
\check{P}_{\pm} = \frac{1}{2} \left( \check{1} \mp \cg \right).
\end{equation}
Here, these projectors satisfy the following relations:
\begin{eqnarray}
\cP_{\pm} \cdot \cP_{\pm} &=& \cP_{\pm}, \label{eq:p1}\\
\cP_+   \cdot \cP_- &=& \cP_- \cdot \cP_+   = \check{0}, \label{eq:p2} \\
\cP_+   + \cP_- &=& \check{1}, \label{eq:p3} \\
\cg &=& \cP_- - \cP_+, \\
	&=& -(2 \cP_+ - \check{1}) = -(\check{1} - 2 \cP_-).
\end{eqnarray}
Substituting these projectors into eq.~(\ref{eq:eilen}), we obtain the equation of projectors
\begin{equation}
{\rm i} \Vec{v}_{\rm F} \cdot \nabla \cP_{\pm} + 
[{\rm i} \omega_n \check{\tau}_3  - \check{\Delta},\cP_{\pm}]=0.
\label{eq:pr}
\end{equation}
Next, we introduce the matrix $\ha$ and $\hb$.
Using $\ha$ and $\hb$, we define $\cP_{\pm}$ as 
%%%%%%%%%%%%
\begin{eqnarray}
\cP_+ &=& \left(
\begin{array}{c}
\hat{1}  \\
-\hb 
\end{array}
\right)
(\hat{1} - \ha \hb)^{-1} (\hat{1}, \ha) 
\label{eq:pp}\\
\cP_- &=& \left(
\begin{array}{c}
- \ha  \\
\hat{1} 
\end{array}
\right)
(\hat{1} - \hb \ha)^{-1} (\hb, \hat{1}),
\label{eq:pp2}
\end{eqnarray}
which satisfies eqs.~(\ref{eq:p1}) and (\ref{eq:p2}).
To satisfy eq.~(\ref{eq:p3}), we define the relation between $\ha$ and $\hb$, which is written as
\begin{equation}
(\hat{1} - \ha \hb)^{-1} \ha = \ha (\hat{1} - \hb \ha)^{-1}.
\label{eq:p4}
\end{equation}
Substituting eqs.~(\ref{eq:pp}) and (\ref{eq:pp2}) into eq.~(\ref{eq:pr}), 
we obtain the matrix Riccati equations.
%%%%%%%%%%%%%%%%%%%%%%%%%%%%%%%%%%%%%%%%%%%%%%
%%%%%%%%%%%%%%%%%%%%%%%%%%%%%%%%%%%%%%%%
\section{Case of ${\rm CePt_3Si}$: Mixed Singlet-Triplet Cooper Pairing and Three-dimensional Fermi Surface}
We calculated the LDOS in the case of ${\rm CePt_3Si}$ in ref.~\citen{nagai}.
${\rm CePt_3Si}$ is the heavy fermion superconductor without inversion symmetry,\cite{Bauer,Bauer2} 
where the singlet pairing and the triplet pairing may be mixed.\cite{frigeri1,frigeri2}
The spin-orbit coupling term exists in the Hamiltonian.
We need to extend our formulation in the body of this paper because the pairing symmetry in this material does not satisfy 
the relation $\hDelta \hDelta^{\dagger} \propto \hat{\sigma}_0$.
The LDOS patterns in ${\rm CePt_3Si}$ are derived from the following discussion.
This discussion can also be applied to the non-unitary triplet superconductors.

Considering the result of numerical calculations by Hayashi \textit{et al}\cite{hayashi_con3},
we assume that the spatial variations of $s$-wave pairing component of pair-potential and that of $p$-wave pairing component
 are the same: $\hat{\Delta} =  \left[ \Psi \hat{\sigma}_0 + \Vec{d}_k \cdot \hat{\Vec{\sigma}} \right]{\rm i} \hat{\sigma}_y 
 = \Delta(\rv) \left[ \tilde{\Psi} \hat{\sigma}_0 - \tilde{k}_y \hat{\sigma}_x + \tilde{k}_x \hat{\sigma}_y \right]{\rm i} \hat{\sigma}_y $,
 with the $s$-wave pairing component $\Psi$.
 Here, the ratio of the singlet to triplet component is defined in the bulk region as $\tilde{\Psi} = \Psi/\Delta$.
 This mixed $s+p$-wave model is proposed for ${\rm CePt_3Si}$\cite{frigeri2}.
The Eilenberger equation which includes the spin-orbit coupling term is given as \cite{hayashi_con,hayashi_con2,Schopohl,Rieck,Choi}
%%%%%%%%%%%%%%%%%%%%
\begin{equation}
{\rm i} \Vec{v}_{\rm F}(\tilde{\kv}) \cdot \Vec{\nabla} \check{g} + 
[{\rm i} \omega_n \check{\tau}_3 - \alpha \check{\Vec{g}_k} \cdot \check{\Vec{S}} - \check{\Delta},\check{g}]=0,
\end{equation}
%%%%%%%%%%%%%%%%%
with
%%%%%%%%%%%%%%%%%
\begin{eqnarray}
\check{g}_k = \left(
\begin{array}{cc}
\Vec{g}_k \hat{\sigma}_0 & 0  \\
0 & \Vec{g}_{-k} \hat{\sigma}_0
\end{array}
\right)
 = \left(
\begin{array}{cc}
\Vec{g}_k \hat{\sigma}_0 & 0 \\
0 & - \Vec{g}_k \hat{\sigma}_0
\end{array}
\right),\\
\Vec{g}_k = \sqrt{\frac{3}{2}}(- \tilde{k}_y, \tilde{k}_x,0), \: \: 
\Check{\Vec{S}} = \left(
\begin{array}{cc}
\Vec{\hat{\sigma}} & 0 \\
0 & \Vec{\hat{\sigma}}^{\rm tr}
\end{array}
\right).
\end{eqnarray}
%%%%%%%%%%%%%%%%%%
We neglect the impurity effect and the vector potential because ${\rm CePt_3Si}$ is a clean and extreme-type-II superconductor.\cite{Bauer}
%%%%%%%%%%%%%%%%%%%%%%%%%%%%
The Riccati equations are written as
\begin{eqnarray}
 \Vec{v}_{\rm F} \cdot \Vec{\nabla} \ha_+ + 2  \omega_n \ha_+ + \ha_+ \hat{\Delta}^{\dagger} \ha_+   -
  \hat{\Delta} & & \nonumber \\ 
   +{\rm i}(\ha_+ \alpha (\Vec{g}_k \cdot \hat{\Vec{\sigma}})^{\rm tr} + \alpha 
 \Vec{g}_k \cdot \hat{\Vec{\sigma}} \ha_+) &=& 0,
 \label{eq:ra}\\
 \Vec{v}_{\rm F} \cdot \Vec{\nabla} \hb_- - 2  \omega_n  \hb_- - \hb_- \hat{\Delta} \hb_- + \hat{\Delta}^{\dagger} & & \nonumber 
 \\
  -{\rm i}(\hb_- 
 \alpha 
 \Vec{g}_k \cdot \hat{\Vec{\sigma}} 
 +  \alpha (\Vec{g}_k \cdot \hat{\Vec{\sigma}})^{\rm tr} \hb_-)&=& 0.
 \label{eq:rb}
\end{eqnarray}
With the Kramer-Pesch approximation as with the means in the previous sections, 
we expand Riccati equations by $y$, $\omega_n$ and $\alpha$.
In the 0-th order, equations~(\ref{eq:ra}) and (\ref{eq:rb}) are the same as ones in the body of this paper.
To obtain the solution of these equations, we assume that the 0-th solutions $\hat{a}_0$ and $\hat{b}_0$ are written as 
\begin{eqnarray}
\hat{a}_0 = \left(
\begin{array}{cc}
a_{11} & a_{12} \\
a_{21} & a_{22}
\end{array}
\right), \: \:
\hat{b}_0 = \left(
\begin{array}{cc}
b_{11} & b_{12} \\
b_{21} & b_{22}
\end{array}
\right),
\end{eqnarray}
and substitute these solutions into the 0-th order equations (\ref{eq:rica0a}) and (\ref{eq:rica0b}) with 
$v_{\rm F} \to v_{\rm F} \sin \chi$.
Solving these equations as simultaneous equations about four variables, we obtain the solutions written as
\begin{eqnarray}
\ha_0= \pm \left(
\begin{array}{cc}
0 & -1 \\
1 & 0
\end{array}
\right),\pm
\left(
\begin{array}{cc}
-{\rm i} \sqrt{\frac{k_-}{k_+}} & 0 \\
0 & -{\rm i} \sqrt{\frac{k_+}{k_-}}
\end{array}
\right), \\
\hb_0= \pm \left(
\begin{array}{cc}
0 & -1 \\
1 & 0
\end{array}
\right),\pm
\left(
\begin{array}{cc}
-{\rm i} \sqrt{\frac{k_+}{k_-}}  & 0 \\
0 & -{\rm i} \sqrt{\frac{k_-}{k_+}} 
\end{array}
\right),
\end{eqnarray}
with $k_{\pm} = \tilde{k}_x \pm i \tilde{k}_y$.
Here, we determine the signs of $\ha_0$ and $\hb_0$ by the fact that the 0-th order Green function 
$\hat{g}_0 = -(\hat{1} + \hat{a}_0 \hat{b}_0)^{-1} (\hat{1} - \hat{a}_0 \hat{b}_0)$ diverges (See \S3.1.2), 
namely $\ha_0 \hb_0 = - \hat{1}$.
We consider the limit of $|\Vec{d}| \rightarrow 0$ and that of $\Psi \rightarrow 0$, where pair-potentials are unitary.
In the limit of $|\Vec{d}| \rightarrow 0$, the pair-potential has the singlet component only.
From eq.~(\ref{eq:ai}), $\ha_0$ is equal to 
$i \hat{\sigma}_y$.
In the limit of $\Psi \rightarrow 0$, the pair-potential has only the triplet component.

%%%%%%%%%%%%%%%%%%%%%%%%%%%%%%%%%%%%
In the first order, eqs.~(\ref{eq:ra}) and (\ref{eq:rb}) are written as
\begin{eqnarray}
 v_{\rm F} \sin \chi \frac{\partial \ha_1}{\partial x}  + 2  \omega_n \ha_0 + \ha_0 \hat{\Delta}^{\dagger}_0 \ha_1 +
 \ha_1 \hat{\Delta}^{\dagger}_0 \ha_0 + \ha_0 \hat{\Delta}^{\dagger}_1 \ha_0 \nonumber \\
  - \hat{\Delta}_1+ {\rm i}(\ha_0 \alpha (\Vec{g}_k \cdot \hat{\Vec{\sigma}})^{\rm tr} + \alpha 
 \Vec{g}_k \cdot \hat{\Vec{\sigma}} \ha_0)= 0,\label{eq:cerica1a}\\
v_{\rm F} \sin \chi \frac{\partial \hb_1}{\partial x}  - 2  \omega_n \hb_0 - \hb_0 \hat{\Delta}_0 \hb_1 
 - \hb_1 \hat{\Delta}_0 \hb_0 -\hb_0 \hat{\Delta}_1 \hb_0 \nonumber \\
 + \hat{\Delta}^{\dagger}_1 -{\rm i}(\hb_0 
 \alpha 
 \Vec{g}_k \cdot \hat{\Vec{\sigma}} 
 +  \alpha (\Vec{g}_k \cdot \hat{\Vec{\sigma}})^{\rm tr} \hb_0)= 0. \: \: \: \: \label{eq:cerica1b}
\end{eqnarray}
To solve the above equations, we consider the solution of the homogeneous equations written as
\begin{eqnarray}
v_{\rm F} \sin \chi \frac{\partial \ha_1}{\partial x}  + \ha_0 \hat{\Delta}^{\dagger}_0 \ha_1 +
 \ha_1 \hat{\Delta}^{\dagger}_0 \ha_0  &=& 0,\label{eq:douji}\\
 v_{\rm F} \sin \chi \frac{\partial \hb_1}{\partial x}   - \hb_0 \hat{\Delta}_0 \hb_1 
 - \hb_1 \hat{\Delta}_0 \hb_0 
 &=& 0.
\end{eqnarray}
We assume the solution:
\begin{equation}
\ha_1 = C \exp \left( \frac{2 \lambda}{v_{\rm F} \sin \chi}\int_0^{|x|} dx' \Delta(x') \right) \hat{A},
\end{equation}
and substitute this into eq.~(\ref{eq:douji}).
Therefore we regard the equations as the equations with variables of vector $\Vec{A} =(a,b,c,d)$ and 
find the eigenvalues $\lambda_i$ ($= -\lambda$) and eigenvectors $\Vec{ A}_i$. 
The same procedure is applied to $\hb_1$. 
With the method of variation of constants [$C \to C(x)$], namely with writing 
 \begin{eqnarray}
  {\hat a}_1 =
\sum_{i=1}^4
 C_i(x) \exp
 \Biggl(
   - \frac{2\lambda_i}{v_{\mathrm F} \sin\chi}
   \int_0^{|x|} d x'
   \Delta(x')
 \Biggr)
 {\hat A}_i,\: \:
\end{eqnarray}
eqs.~(\ref{eq:cerica1a}) and (\ref{eq:cerica1b}) are solved, so that 
we obtain the solutions of eqs.~(\ref{eq:ra}) and (\ref{eq:rb}) up to the first order.
The explicit expressions for the resultant Green function and the LDOS pattern are given in ref.~\citen{nagai}.

%\section{Sample}

%Equations in the appendix will be numbered as (A$\cdot$1), (A$\cdot$2), (A$\cdot$3) \ldots.


\begin{thebibliography}{99} %% The number "99" means that this list has more than nine items.
\bibitem{Hess}H. F. Hess% \textit{et al}.: 
, R. B. Robinson and J. V. Waszczak: 
Phys. Rev. Lett. \textbf{64} (1990) 2711.
\bibitem{Nishimori}H. Nishimori %\textit{et al}.: 
, K. Uchiyama, S. Kaneko, A. Tokura, H. Takeya, K. Hirata and N. Nishida: 
J. Phys. Soc. Jpn. \textbf{73} (2004) 3247.
\bibitem{Gygi91}F. Gygi and M. Schl\" uter: Phys. Rev. B {\bf 43} (1991) 7609.  
\bibitem{hayashi}N. Hayashi, M. Ichioka and K. Machida: 
Phys. Rev. B \textbf{56} (1997) 9052.
\bibitem{Schopohl2} N. Schopohl and K. Maki: Phys. Rev. B \textbf{52} (1995) 490.
\bibitem{ichioka}M. Ichioka% \textit{et al}.: 
, N. Hayashi, N. Enomoto and K. Machida: 
Phys. Rev. B. \textbf{53} (1996) 15316.
\bibitem{Ashida89}M. Ashida, S. Aoyama, J. Hara and K. Nagai: Phys. Rev. B {\bf 40} (1989) 8673.
\bibitem{nagato}Y. Nagato, K. Nagai and J. Hara: J. Low Temp. Phys. {\bf 93} (1993) 33.
\bibitem{Schopohl3} N. Schopohl: cond-mat/9804064.  
\bibitem{eschrig}M. Eschrig: Phys. Rev. B \textbf{61} (2000) 9061.
\bibitem{Shelankov}A. L. Shelankov: J. Low Temp. Phys. {\bf 60} (1985) 29. 

 %%%%%%%%%%%%%%%%%%%%%%%%%%%%%%%%%
\bibitem{Caroli}C. Caroli, P. G. de Gennes and J. Matricon: Phys. Lett. {\bf 9} (1964) 307. 
\bibitem{Kramer}L. Kramer and W. Pesch: Z. Phys. \textbf{269} (1974) 59.
\bibitem{Waxman}D. Waxman: Annals of Physics {\bf 223} (1993) 129. 
\bibitem{nagai} Y. Nagai, Y. Kato and N. Hayashi: J. Phys. Soc. Jpn. {\bf 75} (2006) 043706.
\bibitem{Ueno}Y. Ueno: Master Thesis, University of Tokyo (2003).
\bibitem{Larkin}A. I. Larkin and Yu. N. Ovchinnikov: Zh. ${\rm \acute{E}}$ksp. Teor. Fiz. \textbf{55} 
(1968) 2262, [Sov. Phys. JETP \textbf{28} (1969) 1200].
\bibitem{Eilen}G. Eilenberger: Z. Phys. \textbf{214} (1968) 195.
\bibitem{Serene}J. W. Serene and D. Rainer: Phys. Rep. \textbf{101} (1983) 221.
\bibitem{hayashi_con2} N. Hayashi% \textit{et al}.: 
, K. Wakabayashi, P. A. Frigeri and M. Sigrist: 
Phys. Rev. B. \textbf{73} (2006) 024504.
\bibitem{thesis} This subsection is mainly based on ref. \citen{Ueno}.
\bibitem{Kato}Y. Kato and N. Hayashi: J. Phys. Soc. Jpn. \textbf{70} (2001) 3368.
\bibitem{Kopnin}N. B. Kopnin: J. Low Temp. Phys. \textbf{97} (1994) 157.
%%%%%%%%%%%%%%%%%%%%%%%%%%%55
\bibitem{Maki}K. Maki, P. Thalmeier and H. Won: Phys. Rev. B \textbf{65} (2002) 140502.
\bibitem{Izawa}K. Izawa, K. Kamata, Y. Nakajima, Y. Matsuda, T. Watanabe, M. Nohara, H. Takagi, P. Thalmeier and K. Maki:
Phys. Rev. Lett. \textbf{89} (2002) 137006.
\bibitem{Watanabe}T. Watanabe, M. Nohara, T. Hanaguri and H. Takagi: 
Phys. Rev. Lett. \textbf{92} (2004) 147002.
\bibitem{Ravindran}P. Ravindran, S. Sankaralingam and R. Asokamani: Phys. Rev. B {\bf 52} (1995) 12921.
\bibitem{hayashi4}N. Hayashi, T. Isoshima, M. Ichioka and K. Machida: Phys. Rev. Lett. {\bf 80} (1998) 2921.
\bibitem{eschrig2}M. Eschrig: Ph. D. Thesis, University of Bayreuth (1997).
\bibitem{Bauer}E. Bauer% \textit{et al}.: 
, G. Hilscher, H. Michor, Ch. Paul, E. W. Scheidt, A. Gribanov, Yu. Seropegin, H. No${\rm \ddot{e}}$l, M. Sigrist and P. Rogl: 
Phys. Rev. Lett. {\bf 92} (2004) 027003.
\bibitem{Bauer2}E. Bauer% \textit{et al}.: 
, I. Bonalde and M. Sigrist: 
Low Temp. Phys. \textbf{31} (2005) 748, and references therein.
\bibitem{frigeri1} P. A. Frigeri% \textit{et al.}: 
, D. F. Agterberg, A. Koga and M. Sigrist: 
Phys. Rev. Lett. \textbf{92} (2004) 097001.
\bibitem{frigeri2} P. A. Frigeri% \textit{et al}.: 
, D. F. Agterberg, I. Milat and M. Sigrist: 
cond-mat/0505108.
\bibitem{hayashi_con3} N. Hayashi% \textit{et al}.: 
, Y. Kato, P. A. Frigeri, K. Wakabayashi and M. Sigrist: 
Physica  C {\bf 437-438} (2006) 96.
\bibitem{hayashi_con} N. Hayashi% \textit{et al}.: 
, K. Wakabayashi, P. A. Frigeri and M. Sigrist: 
Phys. Rev. B {\bf 73} (2006) 092508.
\bibitem{Schopohl} N. Schopohl: J. Low Temp. Phys. \textbf{41} (1980) 409.
\bibitem{Rieck} C. T. Rieck% \textit{et al}.: 
, K. Sharnberg and N. Schopohl: 
J. Low Temp. Phys. \textbf{84} (1991) 381.
\bibitem{Choi} C. H. Choi and J. A. Sauls: Phys. Rev. B \textbf{48} (1993) 13684.

%%%%%%%%%%%%%%%%%%%%%%%%%%

\end{thebibliography}
\end{document}